\documentclass[traditabstract]{aa}
\usepackage{graphicx}
\usepackage{txfonts}
\usepackage{longtable,lscape}

\def\eg{\mbox{e.g.}}
\def\etal{\mbox{\rm et al.\ }}

\def\rpro{\mbox{$r$-process}}
\def\spro{\mbox{$s$-process}}
\def\sri{\mbox{$s$-rich}}
\def\spo{\mbox{$s$-poor}}
\def\ncap{\mbox{$n$-capture}}
\def\teff{\mbox{T$_{\rm eff}$}}
\def\logg{\mbox{log~{\it g}}}
\def\vmicro{\mbox{$\xi_{\rm t}$}}
\def\kmsec{\mbox{km~s$^{\rm -1}$}}

\begin{document}

\title{The double sub-giant branch of NGC~6656 (M22): a chemical 
characterization \thanks{Based on data collected at the European 
Southern Observatory with the FLAMES/GIRAFFE spectrograph. }}

\author{
A.\, F.\, Marino\inst{1},
A.\, P.\, Milone\inst{2,3},
C.\, Sneden\inst{4},
M.\, Bergemann\inst{1},
R.\, P.\, Kraft\inst{5},
G.\, Wallerstein\inst{6},
S.\, Cassisi\inst{7},
A.\, Aparicio\inst{2,3},
M.\, Asplund\inst{1,8},
R.\, L.\, Bedin \inst{9},
M.\, Hilker\inst{10},
K.\, Lind\inst{1},
Y.\, Momany\inst{11},
G.\, Piotto\inst{12},
I.\, U.\, Roederer\inst{13},
P.\, B.\, Stetson\inst{14},
M.\, Zoccali\inst{15}
}
\offprints{A.\ F.\ Marino}

\institute{  Max-Planck-Institut f\"{u}r Astrophysik, Karl-Schwarzschild-Str. 1,
             85741 Garching bei M\"{u}nchen, Germany\\
             \email{amarino,mbergema,klind@MPA-Garching.MPG.DE}
             \and 
            Instituto de Astrof\'isica de Canarias, E-38200 La
              Laguna, Tenerife, Canary Islands, Spain\\
             \email{milone,aparicio@iac.es}   
             \and
             Department of Astrophysics, University of La Laguna,
             E-38200 La Laguna, Tenerife, Canary Islands, Spain
           \and
             Department of Astronomy and McDonald Observatory,
            The University of Texas, Austin, TX 78712, USA\\
            \email{chris@verdi.as.utexas.edu}
            \and
             UCO/Lick Observatory, Department of Astronomy and
            Astrophysics, University of California, Santa Cruz,
            CA 95064, USA\\ 
            \email{kraft@ucolick.org}
            \and
            Department of Astronomy, University of Washington,
            Seattle, WA 98195, USA\\ 
            \email{wall@astro.washington.edu}
            \and
           INAF-Osservatorio Astronomico di Teramo, Via
             M. Maggini, 64100 Teramo, Italy\\
             \email{cassisi@oa-teramo.inaf.it}
               \and
              Research School of Astronomy \& Astrophysics, Australian National University,
               Mt Stromlo Observatory, via Cotter Rd, Weston, ACT 2611, Australia\\
            \email{martin@mso.anu.edu.au}
               \and
            INAF-Osservatorio Astronomico di Padova, Vicolo
              dell’Osservatorio 5, 35122 Padova, Italy\\
             \email{luigi.bedin@oapd.inaf.it}
             \and
            European Southern Observatory, Karl-Schwarzschild-Str. 2,
             85748 Garching bei M\"{u}nchen, Germany\\
             \email{mhilker@eso.org} 
             \and
             European Southern Observatory, Alonso de Cordova 3107, Santiago, Chile\\
             \email{ymomany@eso.org}
             \and
            Dipartimento  di   Astronomia,  Universit\`a  di Padova,
             Vicolo dell'Osservatorio 3, Padova, I-35122, Italy\\
             \email{giampaolo.piotto@unipd.it}
             \and
             Carnegie Observatories, 813 Santa Barbara Street,
             Pasadena, CA 91101 USA\\ 
             \email{iur@obs.carnegiescience.edu}
             \and
             Herzberg Institute of Astrophysics,
             National Research Council Canada, 5071 West Saanich Road,
             Victoria, BC V9E 2E7\\
             \email{Peter.Stetson@nrc-cnrc.gc.ca}
           \and
             Pontificia Universidad Cat\'olica de Chile,
             Departmento de Astronom\'ia y Astrofis\'ica, Casilla 306,
             Santiago 22, Chile\\
             \email{mzoccali@astro.puc.cl} 
            }

\date{Received Xxxxx xx, xxxx; accepted Xxxx xx, xxxx}
%__________________________________________________________________
%
\abstract{
We present an abundance analysis of 101 subgiant branch (SGB) stars 
in the globular cluster M22.
Using low resolution FLAMES/GIRAFFE spectra we have determined 
abundances of the neutron-capture 
strontium and barium and the light element carbon.
With these data we explore relationships between the 
observed SGB photometric split in this cluster and two stellar groups 
characterized by different contents of iron, slow neutron-capture process 
(\spro) elements, and the $\alpha$ element calcium, that we previously
discovered in M22's red-giant stars.
We show that the SGB stars correlate in chemical composition
and color-magnitude diagram position: the stars with higher metallicity 
and relative \spro\ abundances define a fainter SGB, while stars with 
lower metallicity and \spro\ content 
reside on a relatively brighter SGB.
This result has implications for the relative ages of the two stellar 
groups of M22.
In particular, it is inconsistent with  a large spread in ages of the 
two SGBs.
By accounting for the chemical content of the two stellar groups,
isochrone fitting of the double SGB suggests that their ages are not
different by more than $\sim$300 Myr. 
}

\keywords{globular clusters: general -- 
          globular clusters: individual: NGC 6656 -- 
          stars: population II -- 
          stars: abundances -- 
          techniques: spectroscopy}

\titlerunning{The split SGB in M22}
\authorrunning{Marino et al.}
\maketitle

%%%%%%%%%%%%%%%%%%%%%%%%%%%%%%%%%%%%%%%
\section{Introduction}
\label{introduction}

Thanks to the large amount of spectroscopic and photometric data assembled 
in the last couple of decades, the assumption that all globular clusters 
(GCs) contain a simple mono-metallic stellar population must be modified.  
Nearly all GCs stars exhibit substantial star-to-star variations in 
light elements, mainly C, N, O, Na, Mg, and Al (\eg, Kraft
  1994\nocite{kra94}; Ram{\'{\i}}rez \& Cohen 2002;
Gratton \etal\ 2004\nocite{gra04}).
These anomalous abundances appear to be present in stars of all
evolutionary states, including convectively unmixed subgiant branch 
(SGB) and main-sequence turnoff stars.  
This argues that many of the variations were in the birth
material of the stars we see today.
The light element abundances have various correlations and
anticorrelations that point unmistakably to hot H-burning ON, NeNa,
and MgAl proton-capture chains.
These cannot be products of the present low mass GC stars, so it is
probable that a fraction of GC stars are made up of material processed 
through higher mass stars that are now compact objects in the GCs.
Thus multiple stellar generations in GCs are needed 
(\eg, Marino \etal\ 2008\nocite{mar08}, Carretta \etal\ 2009\nocite{car09}).

In clusters that we may call {\it normal GCs}, stellar abundances of 
elements heavier than
those affected by H-burning show both intra- and inter-cluster
consistency, and their abundances resemble the halo field 
compositions at similar overall metallicities.
For these GCs a two-generation model is sufficient: a primordial 
generation similar to the field, and stars formed as second
generation(s) enriched in material processed through hot H-burning.
Powerful photometric tools to separate these stellar generations 
along the RGBs include the Johnson $U$ band and specific Str\"omgren indices
(Marino \etal\ 2008\nocite{mar08}; Yong \etal\ 2008\nocite{yon08}).
The accepted cluster evolutionary scenario is that {\it normal GCs} 
have been polluted with hot H-burning products by first generation 
asymptotic giant branch stars (AGB, Ventura \etal\ 2001\nocite{ven01}; 
D'Antona \& Caloi 2004\nocite{dan04}), and/or fast rotating massive stars 
(Decressin \etal\ 2007\nocite{dec07}). Massive binaries have also been
proposed as an alternative source (de Mink \etal\ 2009; 
Vanbeveren, Mennekens \& De Greve 2011\nocite{van11}).  
Stars formed as second generation members were born from the material 
released by these proposed first-generation polluters.

Recent spectroscopic studies have revealed that some GCs have
variations not only in light elements, but also in the bulk
heavy element content.
These clusters, which we will designate
{\it anomalous GCs} (AGCs), have significant metallicity dispersions 
(star-to-star variations in Fe-peak abundances).
GCs that have displayed this anomalous behavior include 
NGC~6656 (M22, Marino \etal\ 2009\nocite{mar09}), 
NGC~2419 (Cohen \etal\ 2010\nocite{coh10}),
Terzan~5 (Ferraro \etal\ 2009\nocite{fer09}), and
NGC~1851 (discovered by Yong \& Grundahl 2008;, and confirmed by
Carretta \etal\ 2010, 2011\nocite{car10,car11}).
All these objects share superficial similarities with the most 
massive GC $\omega$~Centauri, whose huge metallicity variations have 
been known since the 1970s (e.g. Dickens \& Wooley 1967\nocite{dw67};
Freeman \& Rodgers 1975\nocite{fr75}; and more recently Norris \& Da
Costa 1995\nocite{ndc95}; Suntzeff \& Kraft 1996\nocite{sk96}; 
Johnson \& Pilachowski 2010\nocite{jp10}; Marino \etal\ 2011b\nocite{mar11b}). 
Omega~Cen shows a very broad metallicity
distribution that could be consistent with 5-6 groups of stars with
different metallicity, as both spectroscopic and photometric studies
seem to suggest (Johnson \& Pilachowski 2010\nocite{jp10}; Marino
\etal\ 2011b\nocite{mar11b}; Sollima \etal\ 2005\nocite{sol05}; Bellini
\etal\ 2010\nocite{bel10}).
The $\omega$~Cen metallicity spread is so large that it may have a
different origin with respect to the other GCs.
It could be the surviving
nucleus of a dwarf galaxy tidally disrupted by the Milky Way, as
suggested by Bekki \& Norris (2006)\nocite{bek06}.
Differently from the simple {\it normal GCs}, in these objects
successive generation(s) may need to be invoked, with supernovae also
playing a role in the pollution of intra-cluster medium.

In the AGCs NGC~1851 and M22, different groups of stars with different 
slow-process ($s$-process) element abundances have been identified 
(Yong \& Grundahl 2008\nocite{yon08} for NGC~1851, and 
Marino \etal\ 2009, 2011a\nocite{mar09,mar11a} for M22, 
hereafter M09 and M11a respectively).
Multiple stellar groups in M22 and NGC~1851 are also clearly 
manifest by 
a split in their SGB color-magnitude diagram domains, as revealed by 
Hubble Space Telescope ({\it HST}) images (Milone \etal\ 2008\nocite{mil08}; 
M09\nocite{mar09}; Piotto 2009\nocite{pio09}).
The split SGB in these two clusters appears to be related to chemical 
differences observed among their red-giant branch (RGB) stars.

The chemical complexities in M22 RGB stars have been extensively 
studied by M09 and M11a\nocite{mar09,mar11a}.
They show that this cluster hosts two metallicity groups, with mean
abundances $<$[Fe/H]$>$~= $-$1.82 ($\sigma$~= 0.07) and 
$-$1.67 ($\sigma$~= 0.05).
These two metallicity groups are characterized principally 
by different relative contents of the $n$-capture elements that
can be efficiently synthesized in the \spro.
That is, $<$[Y,Zr,Ba,La,Nd/Fe]$>$~= $-$0.01 ($\sigma$~=0.06) in the
lower metallicity group, and $+$0.35 ($\sigma$~=0.06) in the high 
metallicity group.
On the other hand Eu, which predominantly is synthesized in the \rpro, 
exhibits constant relative abundances, within the observational errors:  
$<$[Eu/Fe]$>$~= +0.49 ($\sigma$~=0.05) and +0.42 ($\sigma$~=0.08)
in the lower and higher metallicity groups, respectively.
This clearly indicates that the higher $n$-capture element content in the
higher metallicity M22 RGB stars is due to addition of material produced 
via the \spro.
The two stellar groups were named \sri\ and \spo\ by
M09\nocite{mar09}, and we will follow that convention here. 

For M22, M09\nocite{mar09} also demonstrated 
that stellar models cannot entirely reproduce the size of the 
SGB photometric split by considering
only its metallicity spread. 
They suggested that the origin of the split
could be more complex, and involve also difference in age and/or
variations in the total CNO abundance, as proposed by 
Cassisi \etal\ (2008)\nocite{cas08}
and Ventura \etal\ (2009)\nocite{ven09} for NGC~1851.
This scenario is supported by observational evidence for total C+N+O
variations among RGB stars both in NGC~1851 
(Yong \etal\ 2009\nocite{yon09}), and M22 (M11a). 

Although photometric evidence for the population multiplicity of M22 is 
most clearly evident in the SGB domain, previous detailed abundance 
studies have been carried out only for the brighter RGB and AGB stars.
In this study we eliminate this sample mismatch by performing a 
chemical composition analysis of 101 SGB stars in M22.
The layout of this paper is as follows: \S\ref{data} is an
overview of the data set; \S\ref{analysis} contains a 
description of model atmospheres and abundance derivations;
\S\ref{results} presents the abundance results, that are discussed in
\S\ref{dsgb} and \S\ref{enri}. The findings of this paper are 
summarized in \S\ref{conclusions}.

%%%%%%%%%%%%%%%%%%%%%%%%%%%%%%%%%%%%%%%%%%%%%%%%%%%%%%%%%%%%%%%%%%%%%%%%%%
\section{Observations and data reduction\label{data}}
%%%%%%%%%%%%%%%%%%%%%%%%%%%%%%%%%%%%%%%%%%%%%%%%%%%%%%%%%%%%%%%%%%%%%%%%%%

Basic information for M22 can be found in Harris 
(1996\nocite{har96})\footnote{ 
{The 2010 updated version of the Harris catalog is available at}
{\sf http://www.physics.mcmaster.ca/~harris/mwgc.dat}}.
At a distance of $\sim$3.2~kpc M22 is one of the GCs closest to the Sun. 
It has an half-light radius of 3.36$\arcmin$ and a mass of
$log\frac{M}{M_{\odot}}\sim 5.5$, as listed in 
Mandushev \etal\ (1991)\nocite{man91}.
In this section we consider in turn the photometric and spectroscopic
data that we have employed in this study.

\subsection{The photometric dataset\label{sect_phot}}
 
We first establish that the SGB of M22 is photometrically split in a
manner that mimics the division already established among the RGB
(M11a\nocite{mar11a}). We then consider three distinct
sets of photometry available in the literature, 
in order to investigate the distribution of spectroscopic targets in
the color-magnitude diagram (CMD). 
Ground-based observations were used to analyze the CMD over a wide 
spatial field in the $B$ and $V$ bands, and to estimate the 
atmospheric parameters of the spectroscopic targets.  
In addition, we used ground-based $U$ images available for a smaller 
field, and images taken with the Advanced Camera for Surveys 
on board the Hubble Space Telescope ({\it ACS/HST}; 
Clampin \etal\ 2002\nocite{cla02} and references therein) 
in the F606W and F814W bands to 
make our study of the double SGB
extend % the study of the double SGB 
from the ultraviolet to the infrared spectral regions.

The {\it HST} {\it ACS/WFC} images were obtained under program
GO-10775 (PI Sarajedini). 
The data sets consist of (short exposure, deep exposures) = 
3s, 4$\times$55s in F606W, and 3s, 4$\times$65s in F814W.
We used the photometric catalogs provided by Anderson \etal\
(2008) reduced as described in
Anderson \& King (2006)\nocite{ak06}. 

The ground-based photometric database consists of a total of 533 
individual CCD images taken at different telescopes (see Tab.~\ref{stetson}).
These images were taken as part of P. B. Stetson's program to produce 
a large homogeneous globular cluster database.
The images were reduced and calibrated as described in detail by 
Stetson (2000, 2005\nocite{ste00,ste05}).
The final photometric catalog covers a total field of view of
$\sim$34$\arcmin$$\times$33$\arcmin$ and contains 730,432 entries; 
of these, 604,979 objects had sufficient data to allow calibration in 
at least $V$ and one other filter. 
We used the $B$ and $V$ magnitudes from this catalog.

In addition to this wide field catalog, we used a separate photometric
catalog derived from images collected by the SUperb-Seeing 
Imager (SUSI2) camera, previously mounted on 
ESO-NTT telescope (Tab.~\ref{stetson}).  
The SUSI2 camera was a mosaic of two $2k\times4k$, $0\farcs085$ pixel CCDs, 
where each chip covered a field  of view of $5\farcm5\times2\farcm7$.  
The photometric reduction and calibration of this data-set was presented 
in Momany \etal\ (2004)\nocite{mom04}.
In this catalog, $U$ and $V$ magnitudes are available.

Since we were interested only in target stars with high-accuracy 
photometry,
we included in our analysis only relatively isolated stars with good
values of PSF-fit quality indices and small errors in photometry and
astrometry.
A detailed description of the selection procedures is given in 
Milone \etal\ (2009)\nocite{mil09}. 

M22 has an average reddening $E(B-V)$~=~0.34 
(Harris 1996\nocite{har96}); such a large reddening value is rarely 
uniform over a cluster face.
Corrections for differential reddening applied to the HST GC dataset is 
discussed in detail by Piotto \etal\ (2012)\nocite{pio12} for the case
of M22. 
To account for the color and magnitude differences that differential
reddening produces in the ground-based CMDs, we used the procedure 
described by Milone \etal\ (2011)\nocite{mil11}. 
In brief, we first drew a main-sequence ridge (fiducial) line by putting 
a best-fit spline through the median colors found in successive short 
intervals of magnitude. 
We iterated this step with an outlier sigma clipping. 
Then for each program star, we estimated how much the observed stars in 
its spatial vicinity systematically lie to the red or the blue of the 
fiducial sequence.
This systematic color and magnitude offset, measured along the
reddening line, is indicative of the local differential reddening.

\begin{figure*}
\centering
\includegraphics[width=13.2cm]{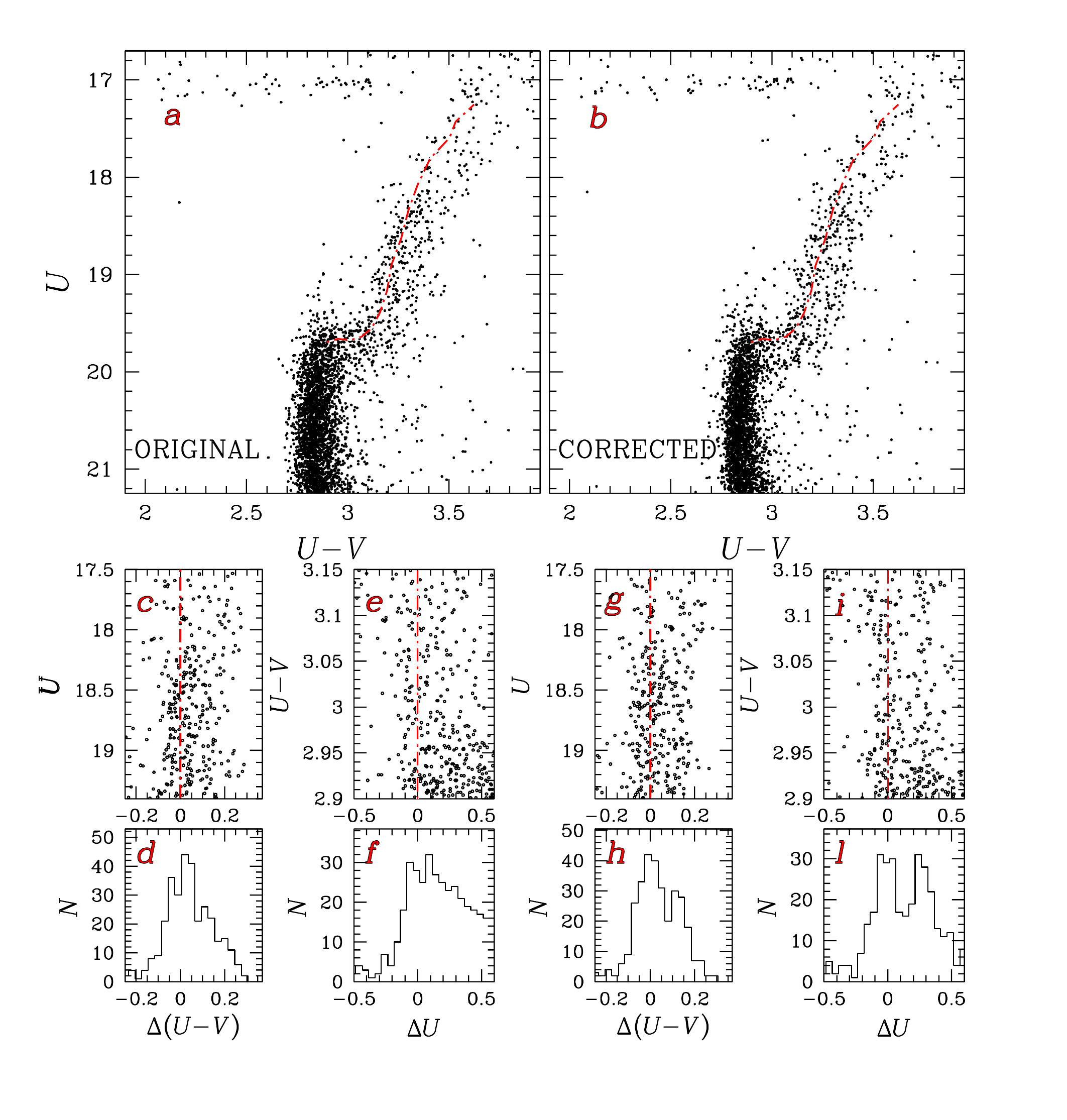}
\caption{Comparison of the $U$ versus $(U-V)$ CMD from NTT
  photometry (Momany et al.\ 2004) before
  (panel \textit{a}) and after (panel \textit{b}) the
  correction for differential (but not for the absolute) reddening. 
  The horizontal branch stars lie at $U$$\sim$17. 
  The $\Delta (U-V)$ distributions of RGB stars with respect to the
  fiducial (red-dotted line on the CMDs) are shown in panels \textit{c}
  and \textit{d} for the non-corrected CMD, and in panels
  \textit{g} and \textit{h} for the corrected one.  
  For SGB stars we show the distributions of the $\Delta U$ magnitude 
  relative to the fiducial shown in the CMDs
  for the non-corrected (panels \textit{e} and \textit{f}) and the
  corrected (panels \textit{i} and \textit{l}) CMDs.   }
\label{t}
\end{figure*}

We corrected for differential reddening all of the {\it HST} and 
ground-based color-magnitude diagrams used in this paper.
As an example, in Fig.~\ref{t}, we compare the original 
(panel \textit{a}) and the corrected $U$-$(U-V)$ CMD 
(panel \textit{b}) of M22 
for the SUSI2@NTT photometry from Momany \etal\ (2004)\nocite{mom04}.   
The choice of this combination of magnitude and color 
is due to its ability to separate photometric sequences at 
different evolutionary stages along the CMD.
The blue and near-$UV$ regimes of cool-star spectra contain 
many CH and CN molecular features.
Stars with different amounts of these elements populate different RGB 
sequences in CMDs constructed by using the $U$-band (Marino \etal\ 2008).  

Inspection of Fig.~\ref{t} shows that after the differential 
reddening correction has been applied, many features of the CMD became 
narrower and more clearly defined.
To better demonstrate the quality of this correction 
we performed some tests represented in the lower panels of Fig.~\ref{t}.
We drew by hand the fiducials for the brighter SGB and bluer RGB population,
and superimposed them to the CMDs as red dashed-dotted lines.
For each RGB star, we calculated the color difference 
$\Delta (U-V)$ from the fiducial at a given $U$ magnitude, while for 
the SGB stars we calculated the difference in the $U$ magnitude at 
a given color.
These color and magnitude differences, as well as histogram plots, have
been represented in the lower panels of Fig.~\ref{t}.
The verticalized $U$ versus $\Delta (U-V)$ diagram is plotted in panels 
\textit{c} and \textit{d} for RGB
stars with $17.5<U<19.4$ by using original and corrected
photometry respectively. 
The corresponding histogram color distributions are shown in panels 
\textit{d} and \textit{h}.  

Panels \textit{e} and \textit{i} of Fig.~\ref{t} show the 
$(U-V)$ versus $\Delta U$ diagram for SGB stars with $2.90<(U-V)<3.15$ 
obtained from original and corrected magnitudes respectively. 
The corresponding histogram color distributions are plotted in panels 
\textit{f} and \textit{l}. 
The better separation of the SGB and RGB sequences in the corrected
diagram suggests that differential reddening has been substantially
removed.
Our differential reddening correction shows that the maximum values of 
$\Delta E(B-V)$ are approximately of 0.1~mag across the face of the
cluster.

Examination of the reddening-corrected $U$ versus $(U-V)$ diagram
(panel {\it (b)} of Fig.~\ref{t})
clearly reveals that the bright SGB is connected to
the blue RGB, while red RGB stars are the progeny of the faint SGB.
We anticipate that this connection between the two SGBs and RGBs would be
further confirmed by our M22 SGB investiagtion, and we will explore
these relationships in \S\ref{dsgb}.

\subsection{The spectroscopic dataset}

Our spectroscopic data consist of a large number of FLAMES/GIRAFFE 
spectra (Pasquini \etal\ 2002, 2003)\nocite{pas02,pas03} 
 observed under the program 085.D-0698A (PI: Marino).
The low resolution LR02 GIRAFFE setup was employed, which covers 
a spectral range of $\sim$600~\AA\ 
from 3964~\AA\ to 4567~\AA, and provides a resolving power 
$R \equiv \lambda/\Delta\lambda \sim$6,400. 
All our target stars were observed in the same FLAMES plate in four 
different exposures of 46 minutes plus one exposure of 26 minutes, for 
a total observing time of 210 minutes.
The typical S/N of the fully reduced combined spectra is $\sim$90-100 at 
the central wavelength of the spectral range.
Data reduction involving bias-subtraction, flat-field correction, 
wavelength-calibration, sky-subtraction,
has been done by using the dedicated pipeline BLDRS v0.5.3\footnote{
See {\sf http://girbld-rs.sourceforge.net}}. 

In total we gathered spectra for 109 candidate M22 SGB stars.
The M22 SGB stars lie in a $(B-V)$ color region ranging from 
$\sim$0.8 to $\sim$1, and extend in $V$ magnitude
from $\sim$17.8 up to $\sim$16.7.
Cluster membership of the stars was established from the radial 
velocities obtained using the IRAF@FXCOR task, which cross-correlates 
the object spectrum with a template. 
For the template we used a synthetic spectrum obtained through the 
spectral synthesis code SPECTRUM 
(Gray \& Corbally 1994\nocite{gra94})\footnote{
See {\sf http://www.phys.appstate.edu/spectrum/spectrum.html}
for more details.}.
This spectrum was computed with a model stellar atmosphere interpolated 
from the Kurucz (1992)\nocite{kur92} grid\footnote{
{\sf http://kurucz.harvard.edu/grids.html}}, adopting parameters 
(\teff, logg, \vmicro, [Fe/H]) = (6000~K, 3.5, 1~\kmsec, $-$1.70).
Observed radial velocities were corrected to the heliocentric
system. 
The observed/template spectrum matches, after the heliocentric 
correction, yielded for the whole sample 
a mean radial velocity of $-$143~$\pm$~1~\kmsec\ ($\sigma$~=~9~\kmsec). 
This value is in reasonable agreement with the values in the literature 
(e.g, $-$148.8~$\pm$~0.8~\kmsec, $\sigma$~=~6.6~\kmsec, 
Peterson \& Cudworth 1994\nocite{pet94};
$-$146.3~$\pm$~0.2~\kmsec, $\sigma$~=~7.8~\kmsec, Harris 1996\nocite{har96}).
Then we rejected individual stars with values deviating by more than 
3$\sigma$ from this average velocity, deeming them to be probable field stars.
After the rejection of these field stars, our sample of {\it bona fide} 
cluster stars is composed of 101 SGBs.

Basic $UBVI$ photometry for the M22 spectroscopically analyzed 
stars is listed in Tab.~\ref{phot_data_tab}. 
In this table we list the coordinates, the original 
$U$, $B$, $V$, and $I$ magnitudes from the ground-based photometry, 
and the differential reddening correction $\Delta~E(B-V)$ applied to each
target.

%%%%%%%%%%%%%%%%%%%%%%%%%%%%%%%%%%%%%%%%%%%%%%%%%%%%%%%%%%%%%%%%%%%%%%%%%%
\section{Data analysis\label{analysis}}
%%%%%%%%%%%%%%%%%%%%%%%%%%%%%%%%%%%%%%%%%%%%%%%%%%%%%%%%%%%%%%%%%%%%%%%%%%

\subsection{Atmospheric parameters\label{atm}}

Chemical abundances were derived from a local thermodynamic
equilibrium (LTE) analysis by using the latest version of the spectral 
analysis code MOOG (Sneden 1973\nocite{sne73}).\footnote{
Available at {\sf http://www.as.utexas.edu/~chris/moog.html}}

Effective temperatures (\teff)\ were estimated by using the 
Casagrande \etal\ (2010)\nocite{cas10} $(B-V)$-\teff\ calibrations 
(based on the ``infrared flux method'') for main sequence and subgiant stars.
Our colors were corrected for the mean M22 reddening
after accounting for differential reddening 
effects as described in Sect.~\ref{sect_phot}.
Indeed, as discussed in Casagrande \etal, accurate reddening corrections 
are crucial in determining \teff\ via the infrared flux method: 
a shift of only $+$0.01 mag in $E(B-V)$ translates into a \teff\ change
of about $+$50~K.
In the case of M22, differential reddening effects are quite large,
and, if left uncorrected, would yield $(B-V)$-based 
\teff\ errors up to $\sim$500 K.
After applying the differential reddening corrections, 
we estimate that our internal color uncertainties are $\approx$0.01-0.015 mag, 
implying internal uncertainties of $\sim$100 K in \teff.
Of course, some stars in our sample,  mainly those in the central crowded 
field of the cluster where the ground-based photometry is not good, 
have larger photometric errors.  
For these stars we expect to have larger errors in the colors that 
translate into larger internal errors in the derived temperatures. 

Surface gravities (\logg) were obtained from the apparent
$V$ magnitudes, the above \teff,\ assuming a mass M~=~0.80M$_{\odot}$, 
bolometric corrections from Alonso \etal\ (1999)\nocite{al99}, and an 
apparent distance modulus of $(m-M)_{V}$~=~13.60 (Harris 1996, 2010 updated).
This value agrees with the one obtained by our best isochrone-fit value,
$(m-M)_{V}$~=~13.64 (Piotto \etal\ 2012\nocite{pio12}).
Gravities derived in this manner
are affected mainly by the adopted distance modulus
and mass, whose variations systematically change the surface gravities. 
An internal variation of $\sim$0.1 in the adopted stellar masses
modifies \logg\ values of $\sim$0.05 dex, with lower gravities for
lower masses.
Such minor excursions in \logg\ do not affect significantly
the abundances derived in this paper (see the error
analysis in Sect.~\ref{abunds}).
The derived values for \teff\ and \logg\ are listed in
Tab.~\ref{abb_data_tab}. 
They span a range of $\sim$900 K and $\sim$0.70, respectively.

Microturbulent velocities cannot be independently determined from our 
spectra or photometry, so we adopted the Gratton, Carretta \& Castelli
(1996)\nocite{gra96} prescription:
\begin{equation}
\vmicro = 2.22-0.322\times\logg
\end{equation}   
However, since all our SGB stars have similar gravities this relation
predicts very similar \vmicro\ values, with a mean 
$<$\vmicro$>$~=~0.97~$\pm$~0.01~\kmsec\ ($\sigma$~=~0.04).
Therefore we assumed a uniform microturbulence  of 1.0 \kmsec\ for all our
targets. 

For the metallicity of our model atmospheres we used in a general way the 
results of 
M09 and M11a\nocite{mar09,mar11a}, derived from
their analysis of high resolution spectra of a large sample of RGB stars. 
Those papers demonstrated that M22 hosts two groups of stars, one
which is \sri\ and one which is \spo, with different 
metallicities 
by a mean [Fe/H] variation of $\sim$0.15 dex.
The different metallicities are accompanied by large difference in \spro\
elemental abundances, with the \sri\ stars having higher
metallicity with respect to the \spo\ stars.
Given the relatively low resolution of our spectra, it is very
difficult
to detect such small metallicity variations in our SGB program stars. 
A difference of 0.15 dex in metallicity does not lead to
significant departures in the relevant model atmosphere  quantities
(such as opacity).
Nevertheless, in our analysis we have accounted for this difference 
by using the following procedure.
({\it i}) First we adopted the mean metallicity of the cluster,
[A/H]~=~$-$1.76 (M09 and M11a), as the metallicity for a given star.
({\it ii}) Then we derived the stars's abundance of Sr, an element
whose abundance in the Solar system is predominantly produced by 
the $s$-process.
({\it iii}) Having then an estimate of the $s$-process content of each 
star, we adopted in its final model atmospheres the mean metallicity 
obtained in our previous work for the \sri\ and \spo\ stars: 
[A/H]~$\approx$~$-$1.67 for the \sri\ stars, and 
[A/H]~$\approx$~$-$1.82 for \spo\ ones (see their Tab.~7).  

\begin{figure}
\centering
\includegraphics[width=9.cm]{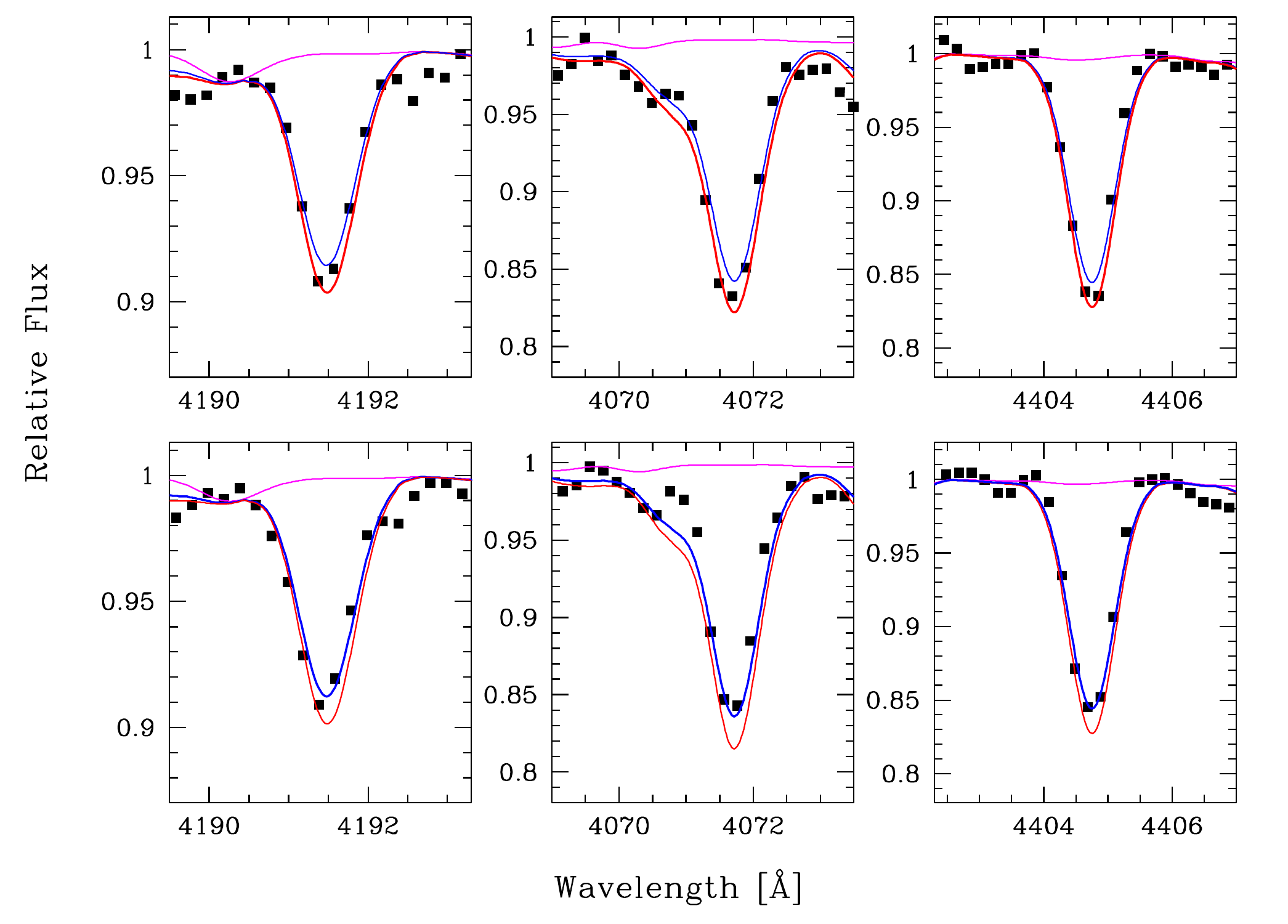}
\caption{\ion{Fe}{I} lines for the combined \spo\ (lower) and \sri\ (upper)
  spectra of Fig.~\ref{spettri}. 
  Superimposed on the observed spectra are synthetic spectra
  corresponding to the [Fe/H]=$-$1.67 (red), [Fe/H]=$-$1.82 (blue),
  and with no Fe (magenta).
}
\label{fe}
\end{figure}

In Fig.~\ref{fe} we show two averaged spectra, covering three \ion{Fe}{I}
lines, obtained from a sample of \sri\ and \spo\ stars (we refer the 
reader to \S\ref{results} for more details). 
On each observed line we have superimposed two synthetic
spectra with appropriate atmospheric parameters, but one with
[Fe/H]~= $-$1.67 (the mean metallicity of \sri\ stars) represented in red, 
and the other with
[Fe/H]~= $-$1.82 (the mean metallicity of \spo\ stars) represented in blue.
The averaged \spo\ star is consistent with having a lower metallicity than 
the \sri\ one, and the level of the Fe difference is similar to the one 
that M09 and M11a found in RGB stars.
However, the line strength differences are small, illustrating the 
difficulty of determining the small difference in metallicity among 
\sri\ and \spo\ SGB stars from our individual spectra.

The contributions to the continuum source
function due to Thomson+Rayleigh scattering effects are small at our
metallicity-temperature-gravity regime for our spectral lines. 
Indeed, we verified that the  differences in Sr, Ba, and C abundances 
obtained with the scattering and non-scattering versions of our
synthetic spectrum code MOOG are negligible.
Therefore, we used the code version that does not take scattering into
account.

In order to keep the present analysis consistent with our previous 
work on M22, we used interpolated model
atmospheres from the grid of Kurucz (1992)\nocite{kur92},
with fixed \teff,\ \logg,\ \vmicro,\ and metallicity.
These models were constructed with inclusion of convective 
overshooting.  
Our trial syntheses suggest that the use of these models 
produces an over-estimation of $\sim$0.10~dex in all abundances
compared with those determined with model atmospheres from 
Castelli \& Kurucz (2004)\nocite{cas04}, which do not assume convective
overshooting.
We emphasize that the metallicity assumptions had little influence 
on establishing the $s$-richness of individual stars; these were
accomplished almost exclusively by the Sr abundances. 
More details on the segregation of stars on basis of
$s$-element content are given in \S\ref{sprosr}.

\begin{figure*}
\centering
\includegraphics[width=11cm]{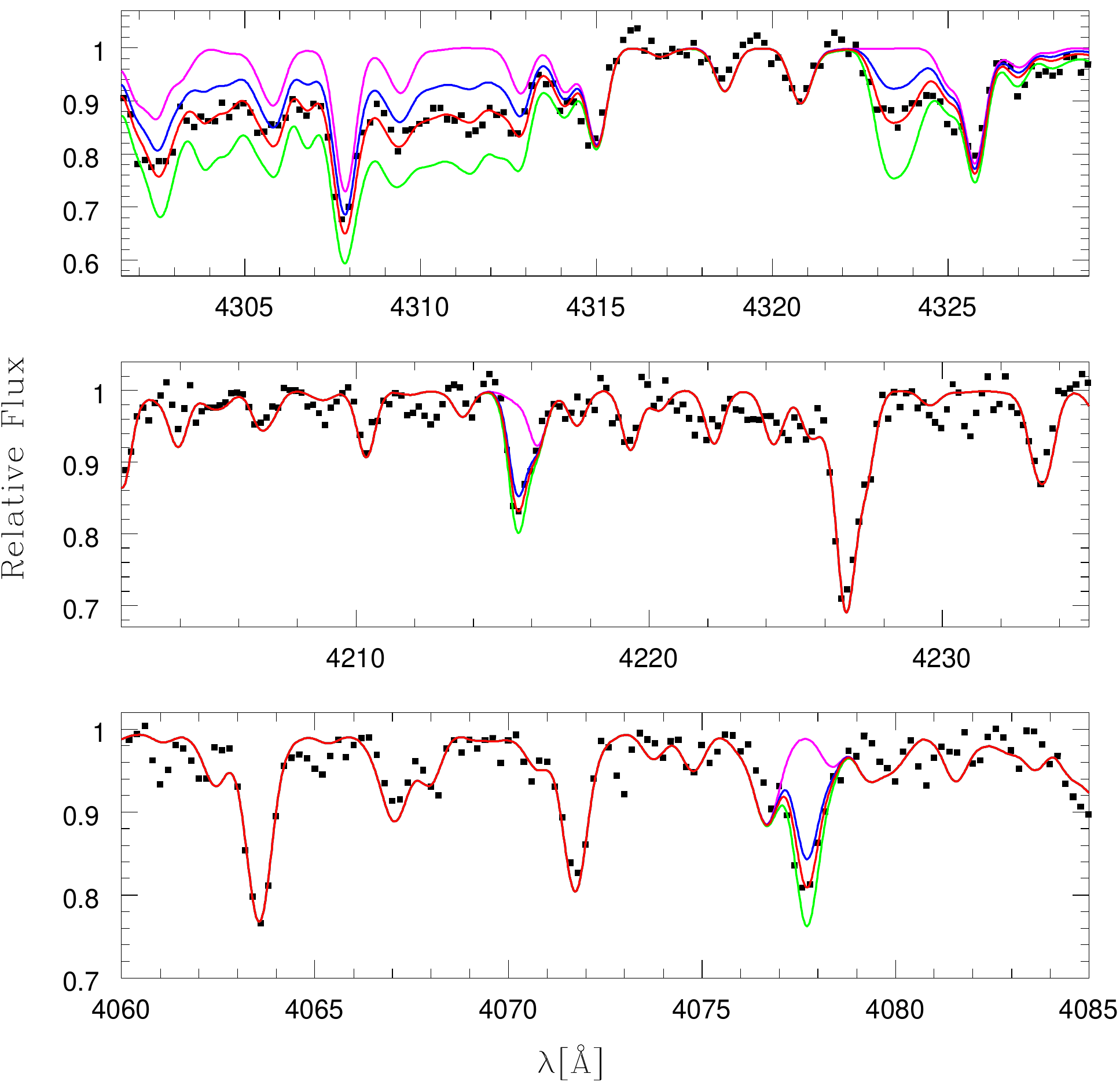}
\caption{Observed and synthetic spectra around the Sr lines at 4077~\AA\
  and 4215~\AA\, and the G band for the \sri\ star \#2505.
  In each panel the points represent the observed spectrum, and the
  continuous lines are the synthesis computed with [Fe/H]=$-$1.76 and
  with different strontium and carbon abundances. 
  The magenta line is the spectrum computed with no contribution
  from \ion{Sr}{II} and C; the red line is the best-fitting synthesis (with
  the abundances given in Tab.~3); and the green and
  blue lines are the syntheses computed with Sr and C  abundances altered by
  $\pm$0.3 dex from the best value.
}
\label{fig1}
\end{figure*}

%%%%%%%%%%%%%%%%%%%%%%%%%%%%%%%%%%%%%%%%%%%%%%%%%%%%%%%%%%%%%%%%%%%%%%%%%%
\subsection{Chemical abundances\label{abunds}}
%%%%%%%%%%%%%%%%%%%%%%%%%%%%%%%%%%%%%%%%%%%%%%%%%%%%%%%%%%%%%%%%%%%%%%%%%%

Using the model atmospheres and analysis code described in \S\ref{atm}, 
we determined abundances for the neutron-capture ($n$-capture) elements 
Sr and Ba and for light element C.

Limited by the relatively low resolution and the small wavelength range
of our spectra, we derived Sr and Ba abundances only from
the strong resonance transitions \ion{Sr}{II} 4077, 4215~\AA, and 
\ion{Ba}{II} 4554~\AA.
Both the Sr lines suffer from many blends with other surrounding
transitions, mostly Fe features and,  in the case of the
\ion{Sr}{II} 4077, also other $n$-capture species (Dy and La). 
Spectral synthesis in the analysis of these lines (and particularly at 
our moderate resolution) is necessary to take these blends into account.
Although the \ion{Ba}{II} 4554~\AA\ is isolated from contaminating
transitions, we computed synthesis also for the Ba spectral line, to 
take its isotopic splitting into account.
The linelists are based on Kurucz line compendium,\footnote{
http://kurucz.harvard.edu/linelists.html}
apart from the Ba transition for which we added hyperfine structure and
isotopic data from Gallagher \etal\ (2010).\nocite{gal10}
For Sr our linelists neglect hyperfine/isotopic splitting;
the wavelength shifts are very small and Sr has one dominant isotope.

The resonance lines of \ion{Sr}{II} and \ion{Ba}{II} are formed 
relatively far out in the atmosphere, 
and, according to our NLTE calculations,
are affected by departures from LTE (see \eg\ Bergemann \& Gehren
  2008 for details on these calculations).
Here, we note that \ion{Sr}{II} and \ion{Ba}{II} are the majority 
ions of their elements in the atmospheres of late-type stars
(as also demonstrated by Short \& Hauschildt 2006\nocite{sho06}),
and the major deviations from LTE are due to the non-equilibrium 
excitation effects in the line transitions. 
In particular, deviation of the line source functions from the Planck 
function leads to the NLTE profile strengthening, thus requiring somewhat
lower abundances to fit observed spectral lines.
For further details
  on the NLTE effects affecting our lines we refer the reader to Bergemann \etal\ (to be
  submitted to A\&A).
Here, we use these NLTE corrections to estimate how much our
abundances could be affected by these effects.
According to our NLTE calculations, we estimated NLTE corrections to range from $-$0.12 to $+$0.04~dex for 
the \ion{Ba}{II} 4554~\AA\ line, and only from $-$0.05 to 0.00~dex for the
two \ion{Sr}{II} lines.

An additional difficulty in the analysis of Ba is that it has five 
major naturally-occurring isotopes whose production fractions in the 
rapid-process (\rpro) and \spro\ are significantly different 
(\eg, Kappeler \etal\ 1989\nocite{kap89}). 
In particular, abundances derived from the \ion{Ba}{II} 4554 \AA\
transition are very sensitive to the adopted $r/s$ fraction 
(e.g.\ Mashonkina \& Zhao 2006\nocite{mas06}; 
Collet \etal\ 2009\nocite{col09}).
This is an issue in the analysis of M22, which hosts stars with
different contribution from the $s$-process material, 
as demonstrated in M09 and M11a.  

The M22 \spo\ stars have \ncap\ abundance distributions that are
compatible with pure \rpro\ material, as shown in 
Roederer \etal\ (2011)\nocite{roe11}.
Since the \sri\ stars should have a larger nucleosynthetic
contribution coming from the $s$-processes (recalling that
within observational errors, Eu is constant in the two M22 $s$-groups), 
a two-step abundance analysis needed to be adopted.
We first determined Ba abundances by adopting a scaled solar-system 
Ba abundance and isotopic fractions (Lodders 2003\nocite{lod03}) for 
all the stars in our sample.
Then the initial Sr syntheses were used to divide the total 
sample into \spo\ and \sri\ groups of stars (as discussed further in 
Sect.~\ref{sprosr}).
Finally we recalculated the Ba abundances of the \spo\ stars, 
assuming a pure \rpro\ isotopic ratio (Arlandini \etal\ 1999\nocite{arl99}).
The re-computed Ba abundances in \spo\ stars are lower by $\sim$0.20~dex 
than the ones obtained in our first abundance estimates.
Of course a similar systematic shift towards lower Ba abundances
is obtained also for the \sri\ stars if a pure \rpro\
isotopic ratio is assumed. However, for the \sri\ stars,
we kept our original Ba abundance values, because
the solar-system isotopic fractions for the \sri\ stars appear to be
a good approximation. 
This is justified since M11a\nocite{mar11a} showed
that the M22 \sri\ RGB stars have a solar-system \ncap-element mix,
i.e., [Y,Zr,Ba,La,Nd/Eu]~$\approx$~0.

Carbon was measured from spectral synthesis of the 
CH ($A^2\Delta-X^2\Pi$) G-band heads near 4314 and 4323~\AA.
The molecular line data employed for CH were provided by B. Plez 
(priv. comm.; some basic details of the linelist
are given in Hill \etal\ 2002\nocite{hil02}).
As an example, we show in Fig.~2 the spectral synthesis of the Sr
lines and the G band for the \sri\ star \#2505 (\teff~=~5923~K,
\logg~=~3.97, [A/H]~=~$-$1.67).
A list of the derived chemical abundances, together with the atmospheric
parameters, is provided in Table~1.

An internal error analysis was accomplished by varying one by one the 
temperature, gravity, metallicity, and microturbulence, and re-determining 
the abundances for three stars spanning the entire range in \teff. 
The parameters were varied by $\Delta$\teff~=~$\pm$100~K,
$\Delta$\logg~=~$\pm$0.2, $\Delta$[Fe/H]~=~$\pm$0.1, and
$\Delta$\vmicro~=~$\pm$0.2, typical uncertainties associated with our 
atmospheric parameters.
The contribution of continuum placement errors was estimated by determining
the change in abundances as the synthetic/observed continuum normalization
was varied: generally this uncertainty added 0.10~dex to the abundances.
The various errors were added in quadrature, resulting in typical 
uncertainties of $\approx$0.15 for the C abundances, $\approx$0.25 for
the individual Sr line abundances, and $\approx$0.22 dex for Ba. 
The standard errors $\sigma$ associated with the mean Sr abundances
obtained from the two available spectral lines, are listed in
Tab.~\ref{abb_data_tab}. The mean of these $\sigma$ values, that is
$0.08\pm0.01$, is an estimate of the error associated with the mean Sr
abundance of each star.
Systematic effects could affect our atmospheric parameters, which would 
lead to systematic abundance differences that could be larger
than those introduced by internal uncertainties.
However, we are interested here only in relative star-to-star chemical 
variations among a set of M22 SGB stars with a restricted parameter range.  
This renders systematic abundance uncertainties unimportant for our
purposes.
Investigation of such systematics is worth pursuing in the future, but
is beyond the aims of our work.

%%%%%%%%%%%%%%%%%%%%%%%%%%%%%%%%%%%%%%%%%%%%%%%%%%%%%%%%%%%%%%%%%%%%%%%%%%
\section{RESULTS\label{results}}
%%%%%%%%%%%%%%%%%%%%%%%%%%%%%%%%%%%%%%%%%%%%%%%%%%%%%%%%%%%%%%%%%%%%%%%%%%

Our results for C, Sr, and Ba abundances in all program stars 
are listed in Tab.~\ref{abb_data_tab}.
These three elements all show a large spread that cannot be entirely 
accounted for by observational errors.
The mean abundances for all the analyzed stars are:
$<$[C/H]$> =-$1.75 ($<$[C/Fe]$> =$0.00,  
$\sigma$~=~0.22, for 100 stars);
$<$[Sr/H]$_{\small \rm NLTE}$$> =-$1.66 
($<$[Sr/Fe]$_{\small \rm NLTE}$$>$ = 0.10, 
$\sigma$~=~0.21, for 101 stars); and
$<$[Ba/H]$_{\small \rm NLTE}$$>=-$1.63  
($<$[Ba/Fe]$_{\small \rm NLTE}$$>$= 0.13,  
$\sigma$~=~0.26, for 100 stars). 

For strontium and barium the mean abundances are those corrected for
NLTE.
In the following we discuss the spreads of each single element.

\subsection{Strontium\label{sprosr}}

In Fig.~\ref{fig2} we plot the LTE and NLTE Sr abundances for our program
stars, both in ``absolute'' log~$\epsilon$ units and relative [Sr/Fe]
ratios.
The large spreads in Sr that we observe here are consistent with 
our findings on RGB stars.
The small NLTE corrections make no significant alterations to these spreads.
No Sr abundances were reported in M09 and M11a\nocite{mar09,mar11a}. 
However, the Sr distribution for our SGB sample is clearly bimodal, similar 
to the M22 distributions of many other \ncap\ elements among RGB stars.

In the following, we use the Sr abundances to divide \sri\ from \spo\ stars. 
Our working hypothesis is to consider the stars having 
log~$\epsilon$($\rm {Sr_{LTE}}$)~$>$~1.40 to be \sri, and the ones with 
log~$\epsilon$($\rm {Sr_{LTE}}$)~$\leq$~1.40 to be \spo.
The Sr distributions, in the left panel of Fig.~\ref{fig2}, illustrate
our chosen selection of \sri\ (red) and \spo\ (blue) stars.
In the right panel we show the two histogram distributions in [Sr/Fe]
of the selected \sri\ and \spo\ stars, constructed by using for the
two groups the mean [Fe/H] values of $-$1.67 and $-$1.82, respectively. 
The adoption of one [Fe/H] value for the \spo\ stars and one for the 
\sri\ stars does not appear to introduce additional spread to 
the distributions of the two groups, and the apparent
larger spreads for the [Sr/Fe] are only due to small binning effects.

Based on our selection, our sample is composed by 56 \spo\ and 45
\sri\ stars. 
The \spo\ stars have log$\epsilon$(Sr)~= 1.09$\pm$0.02
and $\sigma$~=~0.11 ($<$[Sr/Fe]$>$~= $-0.06\pm0.02$),
while the \sri\ ones have log$\epsilon$(Sr)~= 1.59$\pm$0.02 
and $\sigma$~=~0.11 ($<$[Sr/Fe]$>$~= 0.29$\pm$0.02).
The difference in Sr abundances between the two selected groups
is log$\epsilon$(Sr)~= 0.40$\pm$0.03. In order to compare this
result with the ones in M11a, we define the difference in abundance 
ratio for two elements A and B between the \sri\ and \spo\ stars as
$\Delta^{rich}_{poor}$[A/B]~$\equiv$
[A/B]$_{\small \sri}$~$-$~[A/B]$_{\small \spo}$, we obtain
$\Delta^{rich}_{poor}$[Sr/Fe]~$\equiv$ $+$0.35$\pm$0.03.
This difference and the mean [Sr/Fe] values for \sri\ and \spo\ stars 
well agree with the ones of the other $n$-capture elements reported in 
M09 and M11a.
Since in the M22 RGB stars the additional content of $n$-capture elements 
originates from $s$-processes (M11a, Da Costa \& Marino 2009, 
and Roederer \etal\ 2011), the observed Sr increase in a group of SGB stars 
must also have a $s$-process origin.

\subsection{Barium\label{sproba}}

In our previous work on RGB M22 stars we have found a bimodality
in the Ba abundances (M09 and M11a).
However from our analysis on SGB stars, the various Ba abundance 
distributions shown in Fig.~\ref{Ba}
fail to cleanly support the expected bimodality.
This is likely due to the larger uncertainty introduced by the high
sensitivity of the very strong \ion{Ba}{II} 4554~\AA\ line to microturbulent
velocity choices and to assumptions about $r/s$ isotopic fractions,
as discussed in \S~\ref{abunds}.
Here we consider just the isotopic dependence.

In the top panel of Fig.~\ref{Ba} we show the Ba abundances derived 
under assumption that the isotopic ratios are simply those of solar
material.
The \spo\ and \sri\ histograms show very little separation in [Ba/Fe].
In the middle panel we show a more realistic situation by showing LTE
Ba abundances computed using a pure $r$ isotopic ratio for \spo\ stars. 
The mean Ba abundance of the \spo\ stars decreases from 
log$\epsilon$(Ba)~= 0.87$\pm$0.02 
($<$[Ba/Fe]$>_{\rm {\small LTE}}$~= 0.24$\pm$0.02, $\sigma$~= 0.18) to 
log$\epsilon$(Ba)~= 0.67$\pm$0.02 
($<$[Ba/Fe]$>_{\rm {\small LTE}}$~= 0.04$\pm$0.02, $\sigma$~= 0.18).
Finally, in the bottom panel we apply NLTE corrections to the abundances
from the middle panel.
Clearly there is still significant overlap in \spo\ and \sri\ distributions.
However, considering just the NLTE Ba abundances, the mean value for the 
\spo\ stars is log$\epsilon$(Ba)~= 0.60$\pm$0.02 
($<$[Ba/Fe]$>_{\rm {\small NLTE}}$~= $-$0.03$\pm$0.02 ($\sigma$~= 0.16).
This value is very close to the one obtained for RGB stars by analyzing 
transitions in the yellow/red spectral regions,
$<$[Ba/Fe]$>$~= $-$0.05$\pm$0.03 (Tab.~7 of M11a).
The mean Ba content for \sri\ stars, log$\epsilon$(Ba)~= 1.10$\pm$0.03 
($<$[Ba/Fe]$>_{\rm {\small NLTE}}$~= 0.32$\pm$0.03, $\sigma$~= 0.22), also 
agrees with the RGB value, 
$<$[Ba/Fe]$>$~= 0.31$\pm$0.04 (also Tab.~7 from M11a).

\begin{figure}
\centering
\includegraphics[width=9.2cm]{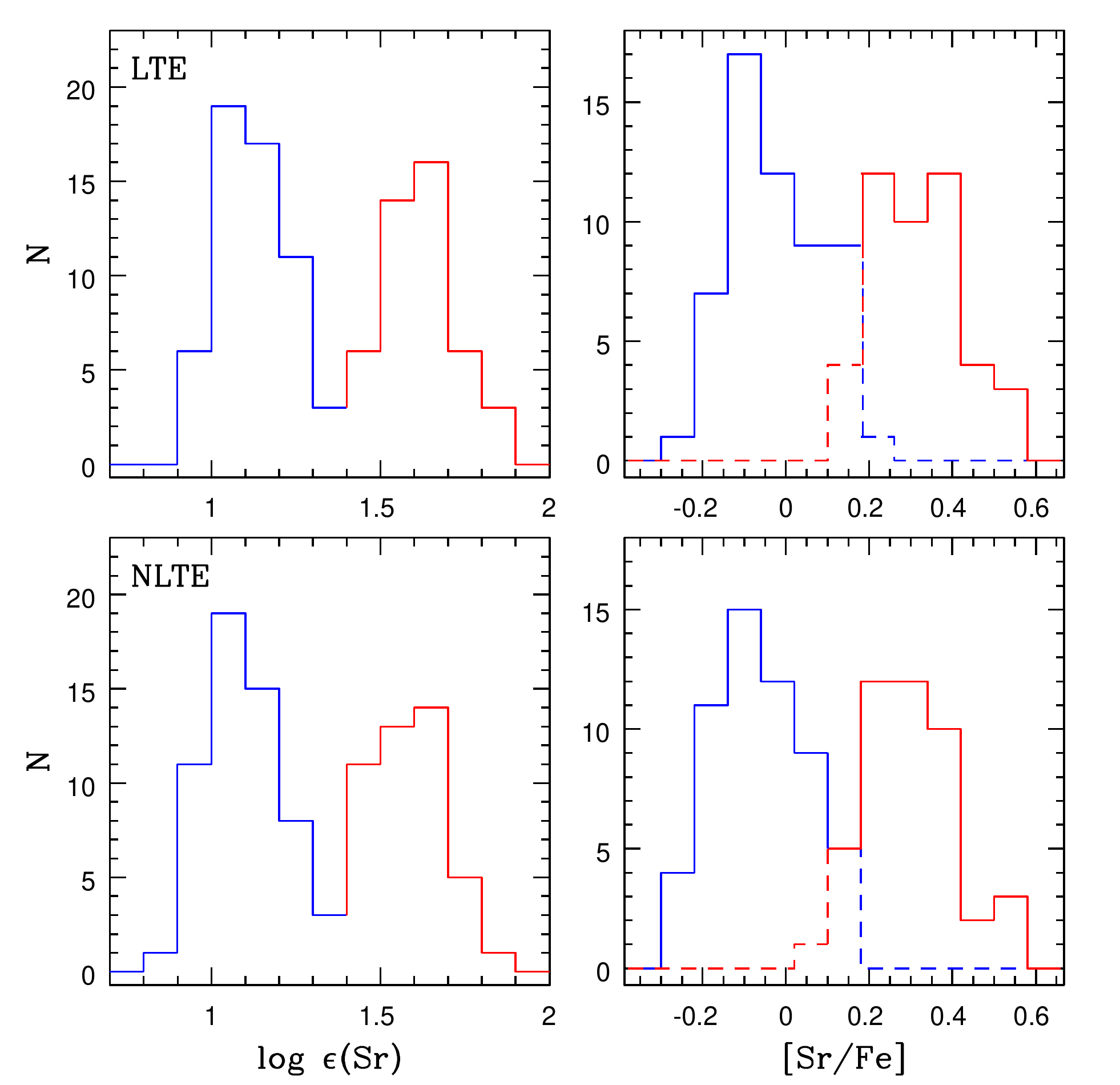}
\caption{{\it Left panels}: Observed distribution of strontium
  abundances in log$\epsilon$(Sr).
{\it Right panels}: Histogram distribution of [Sr/Fe] for the stars
colored in red and blue in the left histograms.
Upper panels show the LTE Sr abundances, lower panels represent the
abundances corrected for NLTE effects.
}
\label{fig2}
\end{figure}

\begin{figure}
\centering
\includegraphics[width=9.4cm]{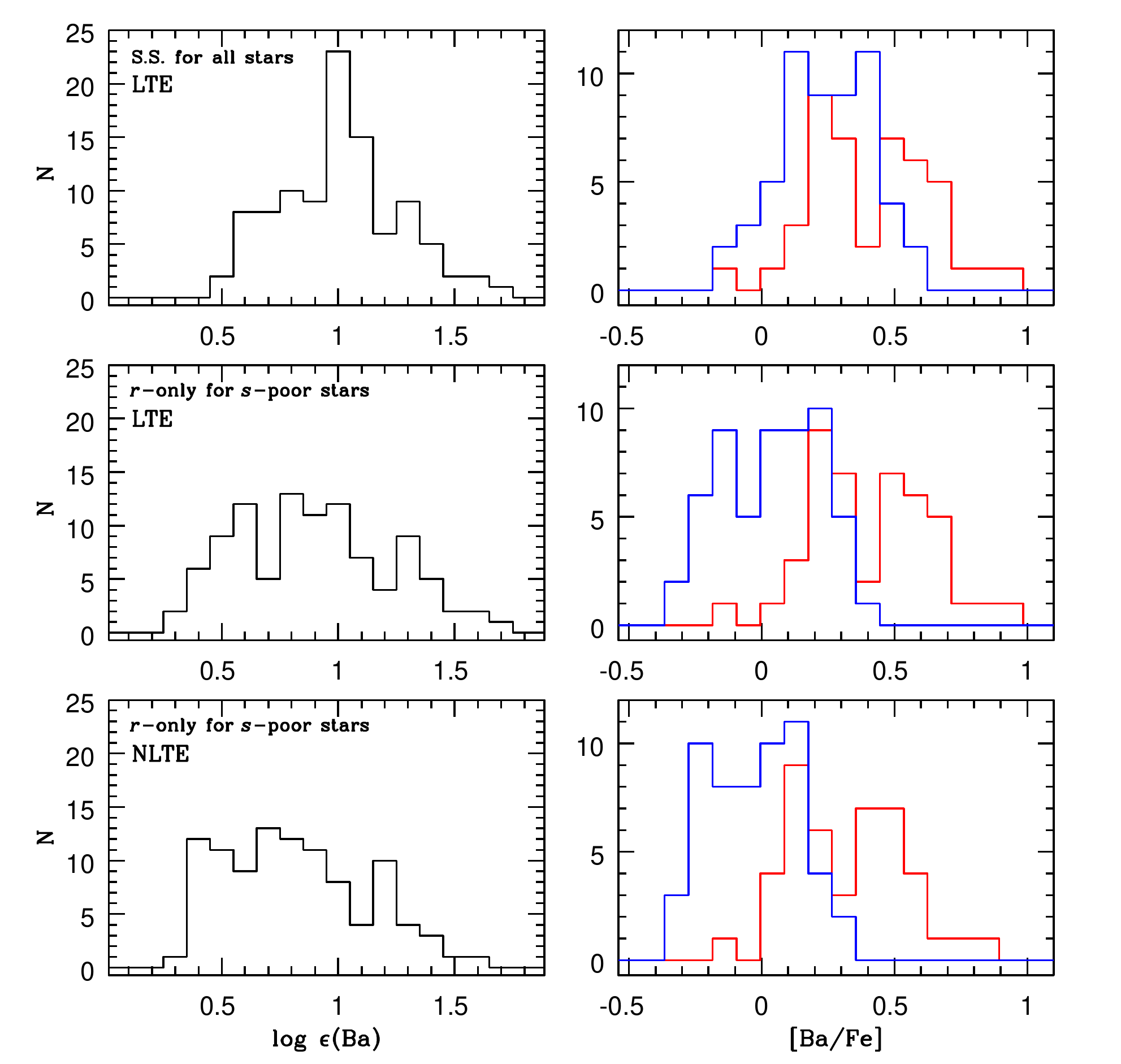}
\caption{Observed distribution of Ba in log($\epsilon$) abundances
  (left panels) and in abundance ratios relative to iron (right panels).
The histogram distributions of [Ba/Fe] for \sri\ and \spo\ stars
selected as in Fig.~\ref{fig2}, have been represented in red and blue
respectively. The upper panels represent Ba abundances with
Solar System (S.S.) isotopic ratios adopted for both \sri\ and \spo\ stars.
In the middle panels an $r$-only isotopic ratio has been applied for
the \spo\ stars. 
The lower panels represent the same abundances represented in the
middle panels corrected for NLTE effects.
}
\label{Ba}
\end{figure}

\begin{figure}
\centering
\includegraphics[width=9.2cm]{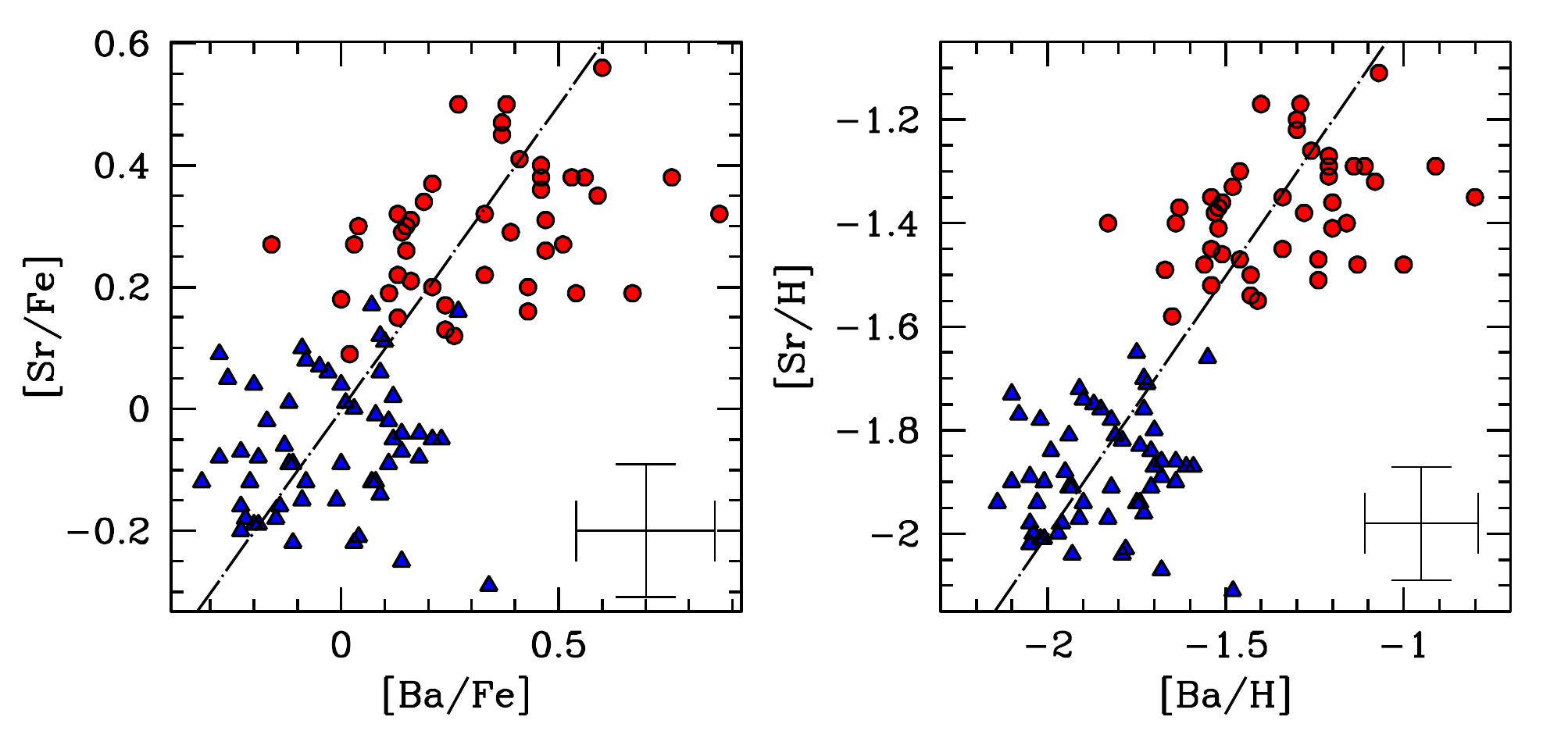}
\caption{[Sr/Fe]  and [Sr/H] as a function of [Ba/Fe] and
    [Ba/H].  \sri\ and \spo\ stars are
  plotted in red circles and blue triangles, respectively. The
  dashed-dotted line represents the perfect agreement.
}
\label{SrBa}
\end{figure}

In comparing the Ba and Sr distributions in Fig.~\ref{fig2} and 
Fig.~\ref{Ba}, the less precise results for Ba are apparent.
The Ba dispersions for both \sri\ and \spo\ stars are much higher than 
the ones for Sr. 
But confidence in the basic results for Ba increases by plotting the
individual Ba and Sr abundances. 
We show this in Fig.~\ref{SrBa}, where our selected \sri\ and \spo\ stars 
have been represented 
 with different symbols in the [Sr/Fe]-[Ba/Fe] and in the
[Sr/H]-[Ba/H] planes.
These elements in the solar system are expected to be equally
sensitive to \spro\ nucleosynthesis (see Tab.~10 in Simmerer
\etal\ 2004\nocite{sim04}). 
In M22 they correlate extremely well on average.
In conclusion, we confirm for SGB stars the \spro\ abundance bimodality
found among the RGB stars in M09 and M11a.

\subsection{Carbon}\label{carbon}

Carbon abundances among RGB stars in M22 have been determined by
M11a. 
They found large spread in C content both among \spo\ and \sri\ 
stars, with the mean C abundance higher for \sri\ stars by 0.35$\pm$0.13.
In addition, in each $s$-group C was found to be anticorrelated with N.
This implies that in each $s$-group, separately, 
a sub-sample of stars that have undergone high-temperature H
burning is present.

We find a large spread of carbon also among the SGB stars.
The \sri\ stars have significantly larger mean 
carbon abundances than those of the \spo\ stars:
log$\epsilon$(C)~= 6.99$\pm$0.03 
($<$[C/Fe]$>$~= $+$0.10$\pm$0.03, $\sigma$~= 0.23), while \spo\ stars have 
log$\epsilon$(C)~= 6.67$\pm$0.02 
($<$[C/Fe]$>$~= $-$0.07$\pm$0.03, $\sigma$~= 0.19).
Thus the abundance difference between the two groups is
$\Delta^{rich}_{poor}$[C/Fe]~= 0.17$\pm$0.04, a more than 
2$\sigma$ difference, consistent with the difference found for 
RGB stars ($\Delta^{rich}_{poor}$[C/Fe]~= 0.35$\pm$0.13, M11a).

\begin{figure*}
\centering
\includegraphics[width=15cm]{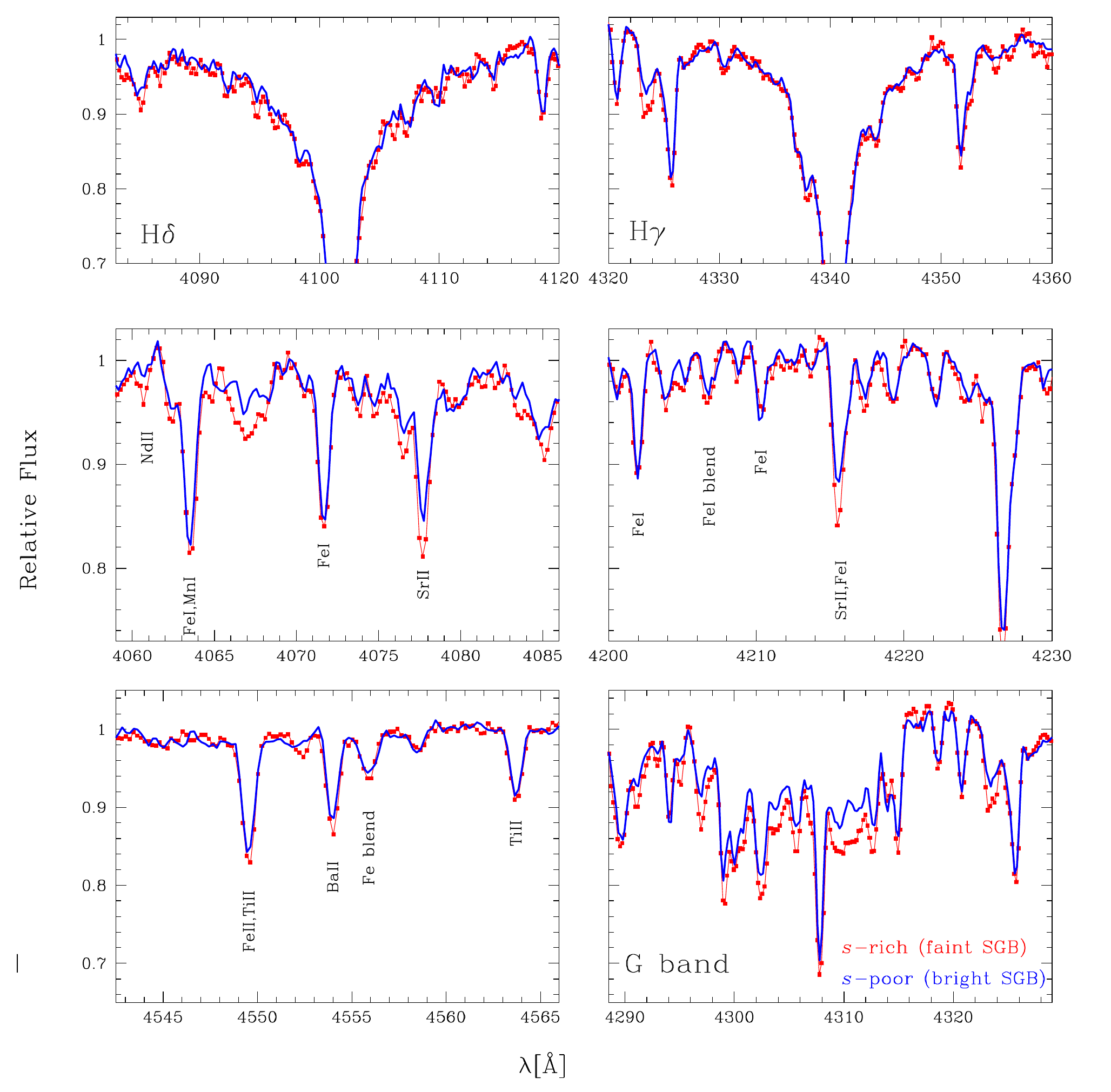}
\caption{Combined \spo\ (blue) and \sri\ (red) spectra constructed by averaging the spectra for eight
 \spo\ stars (\#2544, \#2207, \#2201, \#2801, \#1913, \#735, \#3, \#768) and eight \sri\
 stars (\#1924, \#2659, \#2153, \#2414, \#2404, \#2099, \#2607, \#729) with similar atmospheric parameters.}
\label{spettri}
\end{figure*}

To better visualize our results, we computed an average-\sri\
and an average-\spo\ spectrum by combining stars with very similar
atmospheric parameters. 
The comparison between the two averaged spectra around the spectral 
features of greatest interest is shown in Fig.~\ref{spettri}.
The \sri\ spectrum (in red) clearly shows stronger Sr, Ba CH 
features with respect to the \spo\ one (in blue).
The two available H-lines, shown in the upper panels, show very similar
wings, implying that our atmospheric parameters are reasonably correct
and the averaged stars have similar parameters.

Note that the mean C abundance among RGB stars (M11a) 
is $\sim$0.5 lower than among SGB stars.  
This is likely due to the decrease in [C/Fe] ratios around the RGB bump
($M_{V} \sim 0.5$), and can be explained by the onset of a second 
mixing episode during the red giant evolution of a population II star, 
once the molecular weight barrier established by the retreating 
convective envelope is wiped out by the advancing shell of H-burning 
(e.g.\ Sweigart \& Mengel 1979\nocite{swi79}; 
Charbonnel 1995\nocite{cha95}).  
From that time onward, CN-processed material is able to reach the 
surface layer, where a decrease of C is visible.

\begin{figure}
\centering
\includegraphics[width=9.7cm]{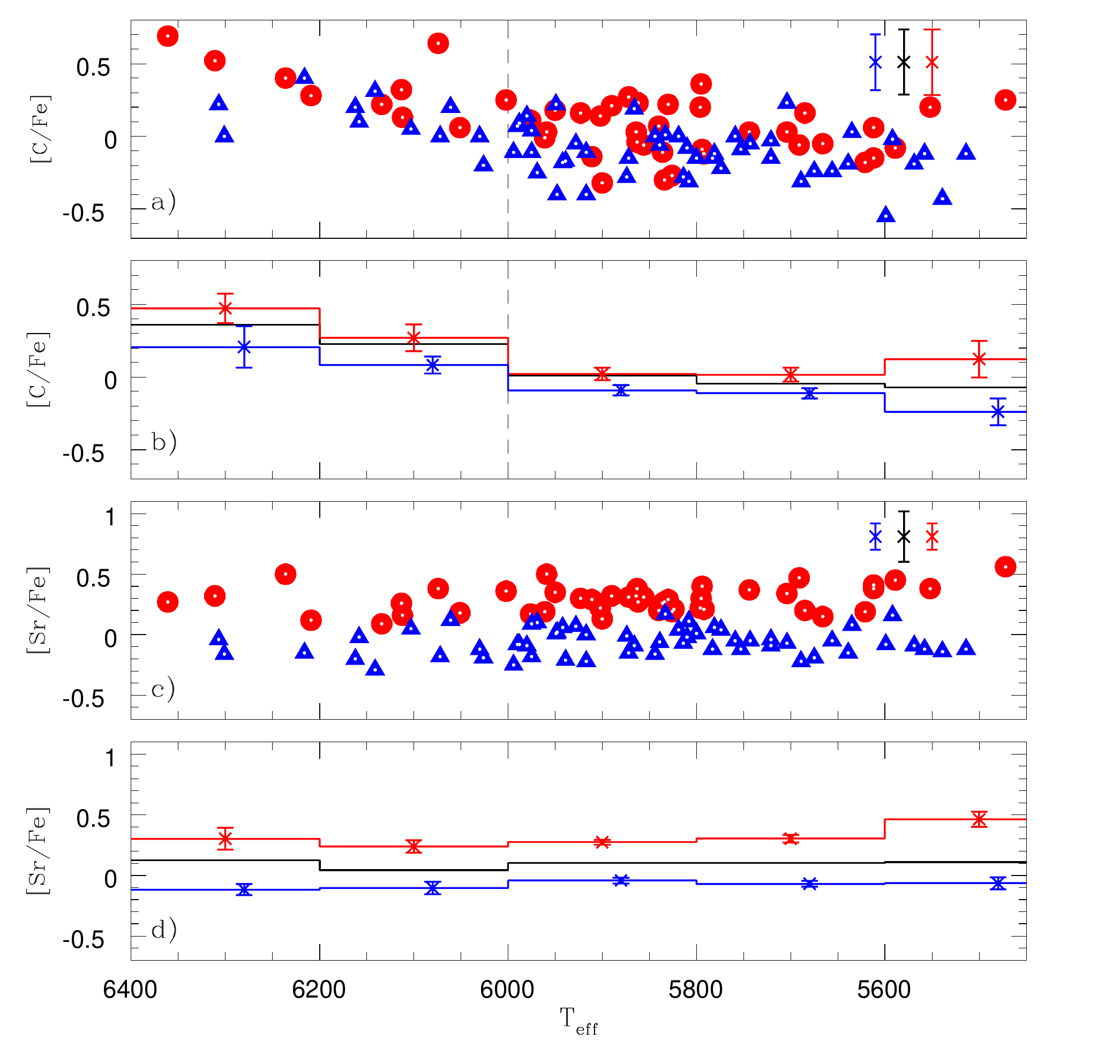}
\caption{C and Sr abundance ratios relative to Fe as a function of
  \teff. In panels {\it a} and {\it c} \spo\ and \sri\ stars are represented in blue triangles and
  red circles, respectively. The observed dispersions for \spo, \sri,
  and the total sample of stars are also shown. 
  In panels {\it b} and {\it d} we plot the mean [C/Fe] and
  [Sr/Fe] values in intervals of 200 K in \teff\ for the two $s$
  groups (red and blue lines), and for the total sample (black line).}
\label{elvsT}
\end{figure}

We also see a trend in [C/Fe] with evolutionary phase, as shown in 
panels {\it (a)} and {\it (b)} of Fig.~\ref{elvsT}, where C abundances 
are plotted as a function of \teff.
As a comparison, Sr abundances represented in panels {\it (c)} and 
{\it (d)} do not show any trend with \teff. 
For clarity, we binned abundances in intervals of 200 K in \teff\
and determined the mean [C/Fe] and [Sr/Fe] abundances for each bin for
\sri\, \spo\ individually, and for the total sample, as represented 
by the red, blue and black histograms respectively.  
To these mean abundances for each bin, we associated
the error bars corresponding to the $\sigma$ divided by $\sqrt{N-1}$, with
N being the number of measurements per bin.
The red, blue and black error bars in the right corner of panels {\it (a)}
and {\it (c)} represent the $\sigma$ values for the 
\sri, \spo, and for the entire sample, respectively.

The carbon rise is particularly pronounced for \teff~$>$ 6000~K.
The reason for this trend is not clear.
For stars with \teff~$>$ 6000~K the uncertainties associated to
the C abundances are surely higher due to the lower line strengths,
and reach values of $\sim$0.20-0.25. These uncertainties could have
led to a systematic over-estimation of C for the hotter stars.
The C rise could also likely be due to our 1D approximation for 
model atmospheres.
A more appropriate analysis of molecular bands should
take into account 3D effects that strongly depend on temperature
(Collet \etal\ 2007\nocite{col07}).
However, since this effect appears to be not significantly 
different between \sri\ and \spo\ stars, it does not affect 
our differential abundance analysis between \sri\ and \spo\ 
stars in M22.

\begin{figure}
\centering
\includegraphics[width=9.3cm]{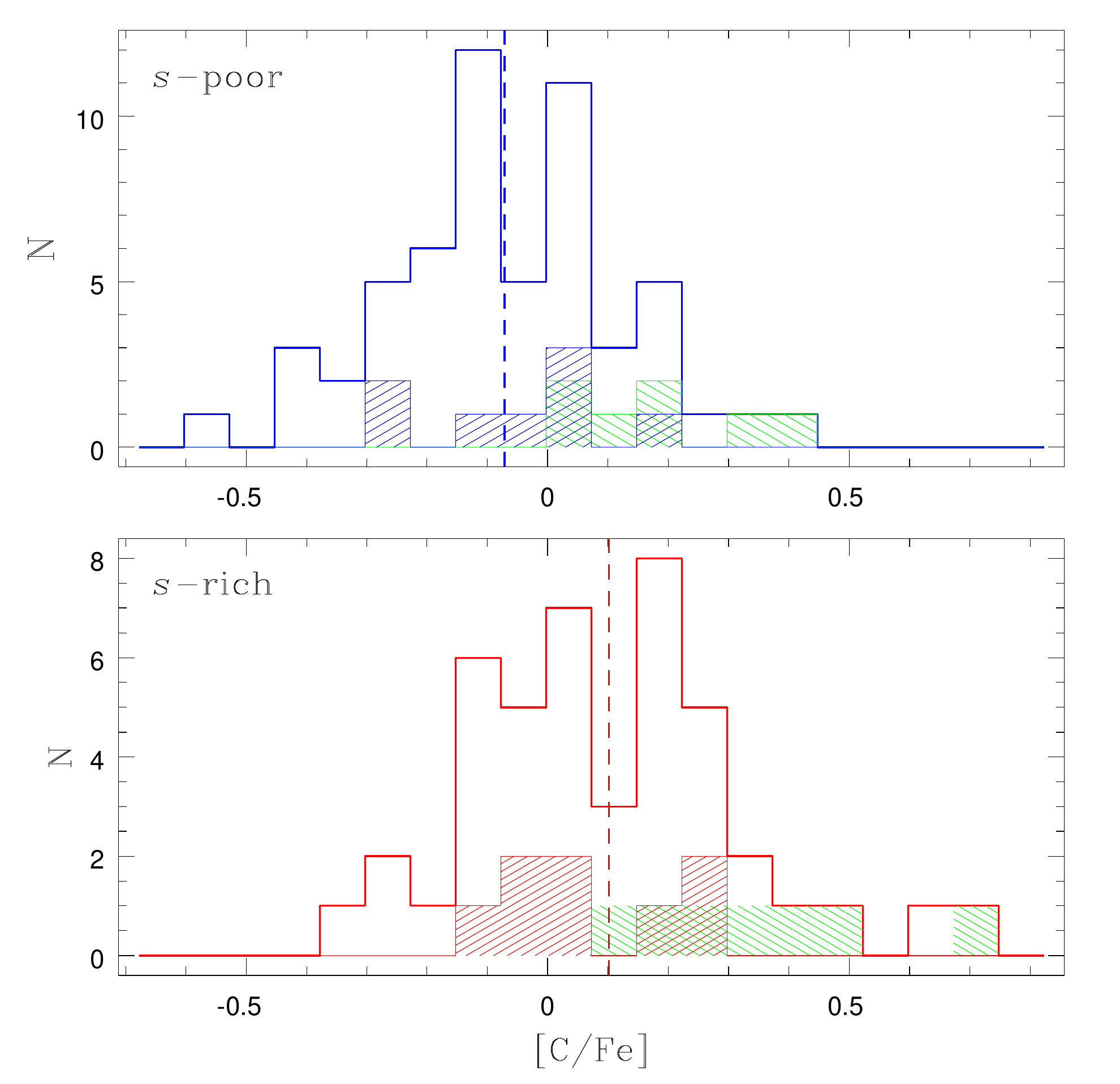}
\caption{Carbon abundance relative to Fe distributions for the \spo\ (upper panel)
  and \sri\ (lower panel) stars. In each panel the dashed line
  represents the mean [C/Fe] abundance. The location of the stars selected for
  constructing the average \sri\ and \spo\ spectrum (see
  Fig.~\ref{spettri}) has been indicated with dashed-red, and blue
  histograms for \sri\ and \spo\ stars respectively.
The dashed green histogram represents the location of stars with
\teff$>$6000 K.
}
\label{Chisto}
\end{figure}

We have already established that the mean value of [C/Fe] is 
substantially higher for the \sri\ stars than for the \spo\ stars 
over the entire range of SGB \teff.
In Fig.~\ref{Chisto} the C distribution is represented for 
the \spo\ stars (upper panel) and \sri\ stars (lower panel).
The shaded blue and red histograms represent the stars used to 
construct the average \sri\ and \spo\ spectra, while the 
green ones represent stars with \teff~$>$ 6000~K (which show 
a rise in C with temperature).
The distribution of carbon for the two groups of stars 
suggests the presence of an intrinsic dispersion in C among 
both \sri\ and \spo\ stars. 
Indeed, both \spo\ and \sri\ stars 
show intrinsic variations of C, N, O, and Na with the presence 
of N-C and Na-O anticorrelations, as revealed in our high-resolution
spectroscopic study on RGBs (M11a). 
The [C/Fe] abundance dispersion for the \sri\ and the \spo\ 
groups here ($\sigma$ is 0.23 and 0.19 dex respectively) are 
marginally larger than the measurement error, which is 
typically 0.15 dex, and indicate the presence of an 
intrinsic [C/Fe] spread.

%%%%%%%%%%%%%%%%%%%%%%%%%%%%%%%%%%%%%%%%%%%%%%%%%%%%%%%%%%%%%%%%%%%%%%%%%%
\section{The double sub-giant branch of M22\label{dsgb}}
%%%%%%%%%%%%%%%%%%%%%%%%%%%%%%%%%%%%%%%%%%%%%%%%%%%%%%%%%%%%%%%%%%%%%%%%%%

M22 is among the GCs showing a double SGB (Piotto 2009, 2012; 
M09\nocite{pio09, pio12,mar09}).  
In the cluster center the bright SGB component is made up
of about 65\% of SGB stars while the remaining $\sim$35\% of stars
defines the fainter SGB component. 
The multi-wavelength study of M22 from Piotto \etal\ (2012)\nocite{pio12} 
reveals that the SGB bimodality is visible in all the bands, from the far
ultraviolet (the F275W HST/WFC3 filter) up to the near infrared 
(the F814W HST/WFC3 filter).
These authors also found that the average magnitude
difference between the bright SGB and the faint SGB is almost the same
at different wavelengths, suggesting that the split reflects internal
structural properties of the stars, and not only the surface composition.

The RGB of M22 is also made up of two main components, but the RGB split
is visible only when appropriate photometric bands are used like the
$m_{1}$ and the $hk$ Str\"omgren indices (Richter \etal\ 1999\nocite{ric99}, 
Lee \etal\ 2009\nocite{lee09}) or the $U$ band (see Fig. 1).
While the SGB split is consistent with two stellar
groups with either an age difference of $\sim$1-2~Gyrs, or a difference 
in their overall C+N+O content, the 
occurrence of simultaneous RGB and SGB
bimodality observed in the $U-V$ color seems to rule out the first
hypothesis of a significant age difference (as suggested by  Sbordone 
\etal\ 2011\nocite{sbo11} for the case of NGC~1851).

\begin{figure*}
\centering
\includegraphics[width=15cm]{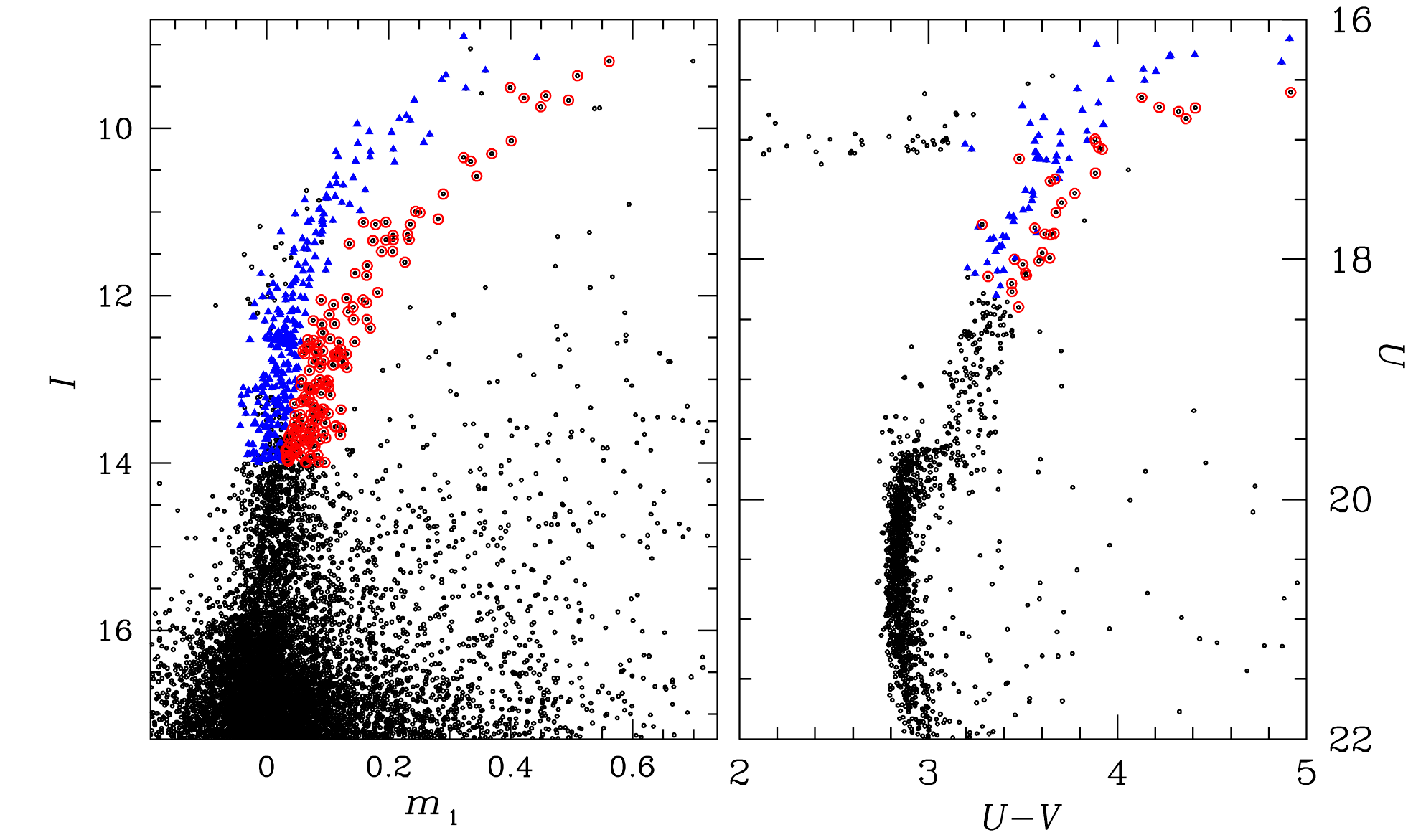}
\caption{{\it Left panel}: $I$ versus the Str\"omgren index
  $m_{1}$. Stars belonging to the two RGBs have been represented in
  red and blue colours. {\it Right panel}: Stars selected in the
  double RGB of the $I$-$m_{1}$ diagram, have been represented in the $U$-($U-B$) CMD. }
\label{conf}
\end{figure*}

M11a have shown that the blue and the red RGB sequences appearing in the 
$I$ versus $m_{1}$ diagram in left panel of Fig.~\ref{conf} are made 
of \spo\ (iron/CNO-poor) and \sri\ (iron/CNO-rich) stars, respectively.  
This CMD is a reproduction of
the same $I$ versus $m_{1}$ diagram  of Fig.~19 in
M11a, but now we have color-coded in red and blue the stars 
photometrically belonging to the two RGBs. 
The same color codes are used to plot these selected stars that are in
common with the SUSI photometry, represented in the
 $U$-$(U-V)$ CMD (right panel of Fig.~\ref{conf}). 
In this latter CMD the two RGBs are clearly connected to the two SGBs, 
providing photometric evidence that the bright SGB and the faint 
SGB are the sub-giant counterparts of the \spo\
and \sri\ RGB, respectively.

Here we can provide direct evidence of the RGBs-SGBs connection,
already clear from the CMD inspection, by matching
our spectroscopic data on M22 SGB stars with the CMDs 
where the split is more clearly visible.
We remind the reader that our photometry has been corrected 
for differential reddening effects applying the procedure described 
in Milone \etal\ (2011, see \S~\ref{data}).

\begin{figure*}
\centering
\includegraphics[width=18cm]{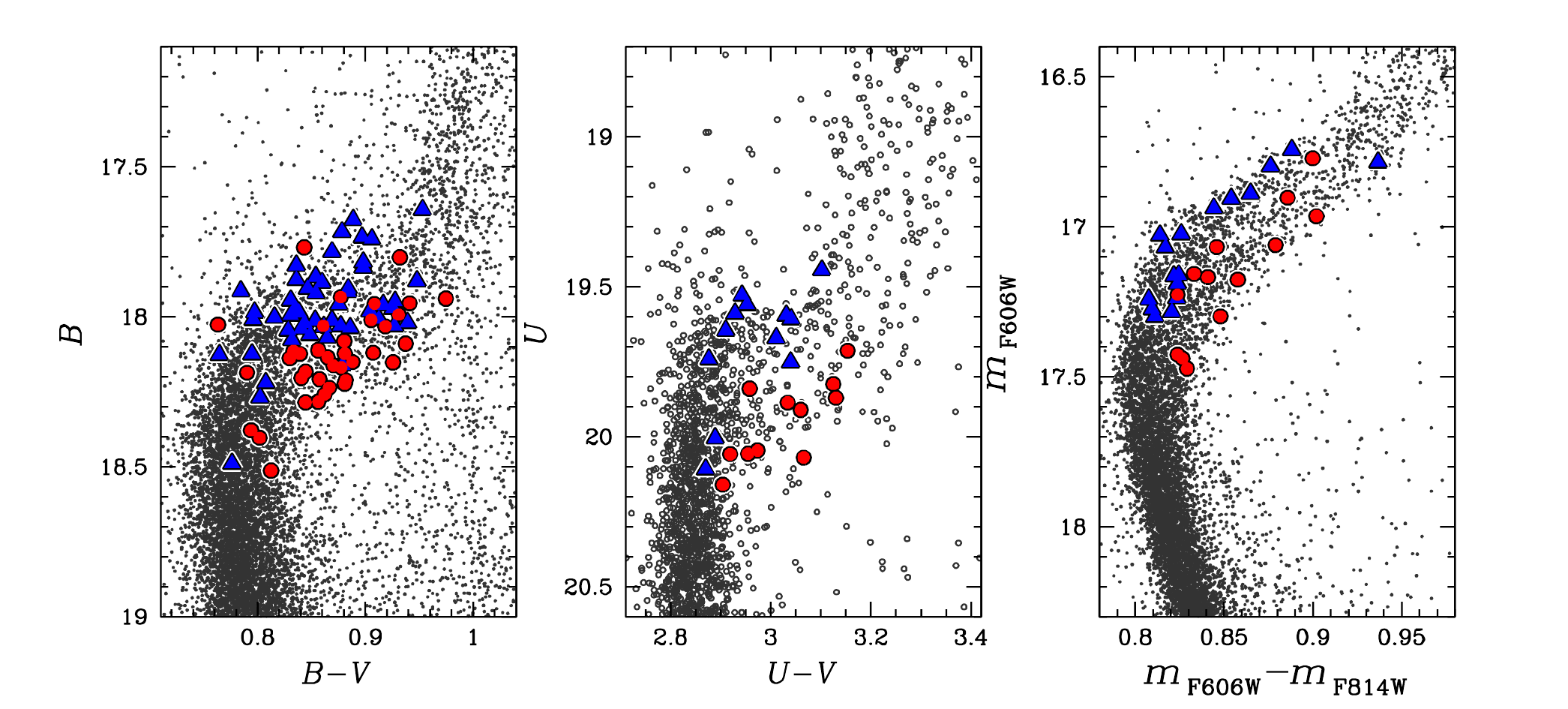}
\caption{M22 double SGB in $B$-($B-V$), $U$-($U-V$), and in the
  $m_{\rm F606W}$-($m_{\rm F606W}-m_{\rm F814W}$) CMDs, with
  the \sri\ and \spo\ stars superimposed.}
\label{sgb}
\end{figure*}

The ground-based and HST CMDs are shown in Fig.~\ref{sgb}, with 
our \sri\ and \spo\ stars superimposed. 
The left panel shows a ground-based $B$-($B-V$) CMD for stars
distributed in a large field (34$\arcmin$$\times$33$\arcmin$) of M22
where we have plotted relatively isolated, unsaturated stars
with good values of the PSF-fitting quality index and small $\sigma$ errors
in photometry and astrometry (see  Sect.~\ref{sect_phot}). 
The $U$-$(U-V)$ CMD in the middle panel is from SUSI2 ground-based 
photometry but covers a small field (two chips of 
2.7$\arcmin$$\times$5.5$\arcmin$) in the outskirt of the cluster. 
The right panel contains the {\it ACS/HST} CMD representing stars
lying in the most central field (3$\arcmin$$\times$3$\arcmin$) of the
cluster.
The double SGB of M22 is clear in all these photometric systems.
By coupling our spectroscpic results with these CMDs, it turns out that
the \sri\ stars occupy the fainter SGB, while the \spo\ ones lie on
the upper brighter SGB.
All the CMDs of Fig.~\ref{sgb} suggest the same:
the \sri\ and \spo\ stars segregate along two different branches
on the M22 SGB, as predicted by M09.
The fact that both the fainter SGB and the redder RGB are made of \sri\ 
stars while \spo\ stars are located on the brighter SGB and the bluer RGB 
further confirms the connection between the two RGBs and SGBs.

Our results suggest that: {\it (i)} the two SGBs are populated by
stars with different $s$-process element content;  {\it (ii)} the
two SGBs "evolve" to the two sequences on the RGB observed in
various M22 CMDs and populated by stars with different $s$-process
elements, overall metallicity, overall C+N+O, and slightly different
Ca, as demonstrated in M11a.
Unfortunately, given the moderate resolution of our SGB spectra, we
cannot distinguish for SGBs stars the small differences in [Fe/H] and
[Ca/Fe] between \sri\ and \spo\ stars as found from high resolution
spectroscopy on RGB. As found for RGB stars, the SGB \sri\ and \spo\
stars have slightly different C abundances, but for SGB stars we have
no information on N and O.
However, given that we have demonstrated that the two SGBs, in the same way
as the two RGBs, have different
$s$-process content and C, and that the sequences are photometrically
linked, we can fairly extend the results on the two RGBs to the two
SGBs for those elements not studied in the present work. 

In the light of our results and previous results on the RGB stars by
M09 and M11a, we can, for the first time, fully
characterize the two SGBs of M22 in terms of chemical composition, as
follows:
\begin{itemize}
\item faint-SGB: \sri\ stars, with higher metallicity, Ca and enhanced
 C+N+O abundance;
\item bright-SGB: \spo\ stars, with lower metallicity, Ca and unenhanced
 C+N+O abundance;
\end{itemize}

The chemical properties for the two SGBs are summarized in Tab.~4.
For the present discussion we consider calcium and the C+N+O sum as
part of the overall metallicity.
Thus we take those values determined from RGB high-resolution spectroscopy 
in M11a, and extend them to the two SGBs with different $s$-process content. 
As discussed in \S\ref{carbon}, elements like C are highly affected by
evolutionary effects that change the surface abundances for stars at
different evolutionary stages, however the total C+N+O is not affected by
these effects.

Theoretical isochrones can reproduce  SGB 
sequences with different luminosities by assuming
an age difference among the two SGB populations.
Under this scenario, in M22 the fainter-SGB would be populated 
by younger stars and the brighter SGB by older ones.
If age is assumed to be the lone factor responsible for
the M22 SGB split, our isochrones can reproduce the observed 
separation in magnitude with an age difference of $\sim$1~Gyr 
between the two SGB populations.
Of course, this is true only in the case of two stellar populations
with identical chemical properties.

However, our results show that there are chemical differences among
the two SGBs and hence the scenario of a simple age difference 
cannot work for this cluster.  
In particular, the overall C+N+O abundance has a strong impact on 
isochrones at SGB luminosities.
This has fundamental consequences for GC age dating, as demonstrated 
in recent literature by Cassisi \etal\ (2008)\nocite{cas08} and
D'Antona \etal\ (2009)\nocite{dan09}.
For NGC~1851 Cassisi \etal\ (2008)\nocite{cas08} and 
Ventura \etal\ (2009)\nocite{ven09} suggested that the
fainter SGB stars could be younger by a few hundred Myrs, if
enhanced in the total C+N+O content. 
Following this scenario large age differences among
the two SGBs could be ruled out.

\begin{figure}
\centering
\includegraphics[width=9cm]{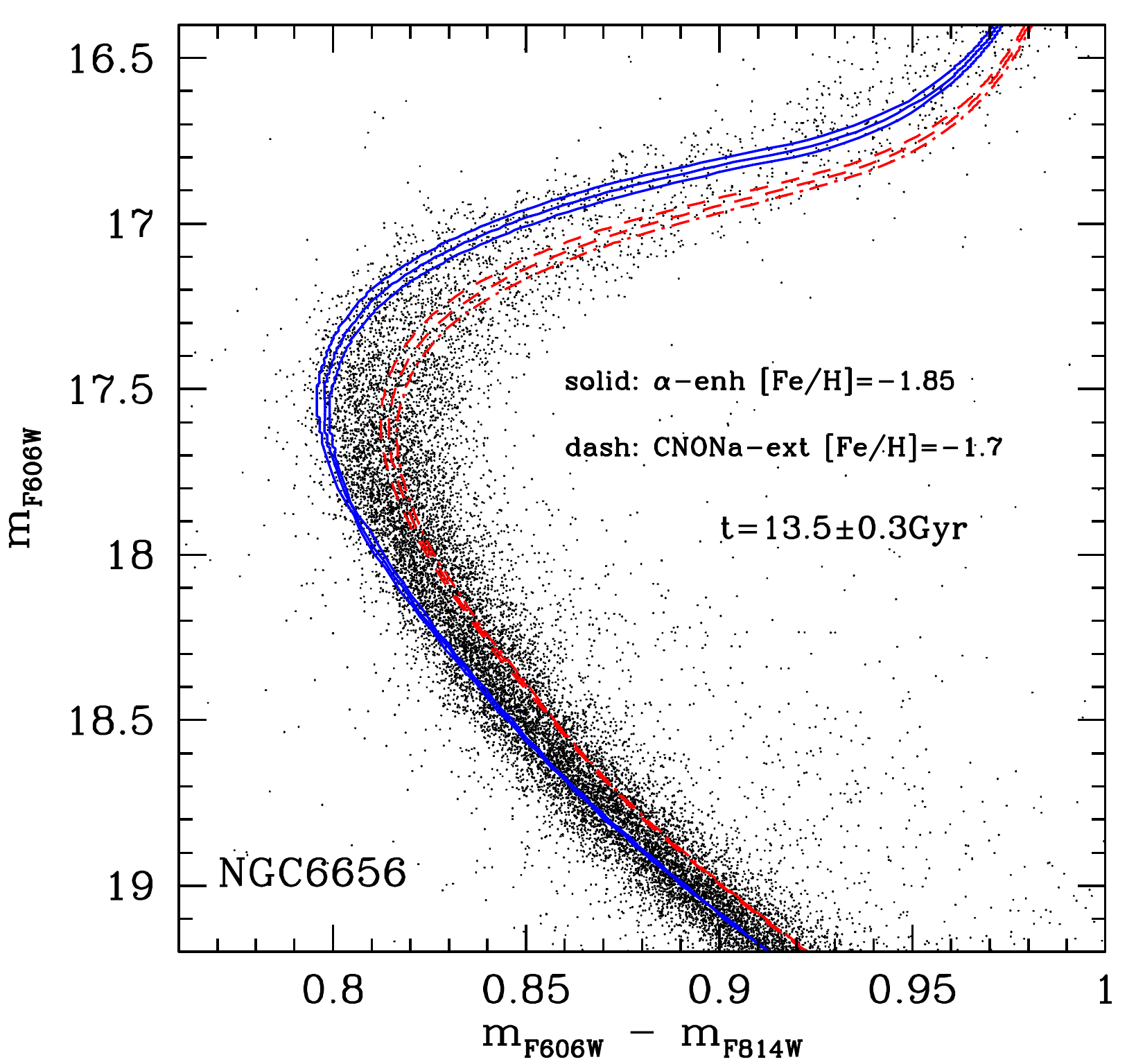}
\caption{Isochrones computed with the observed CNO and metallicity of
  the two $s$ groups,
  and the same age (13.5 Gyr), superimposed to the
M22 SGBs. The red and blue 
tracks are the CNO-higher (\sri), and
CNO-lower (\spo) best-fitting isochrones respectively. 
For both we computed a couple of isochrones with the same chemistry,
but with age varied of $\pm$300 Myrs (dotted tracks).} 
\label{iso}
\end{figure}

In M22 the SGB \sri\ stars are distributed along the faint SGB, indicating
that the faint SGB is composed of stars enriched in \spro\ elements,
and additionally in the total CNO and metallicity,
as suggested by the our previous study on RGB stars.
Here, we use isochrones interpolated in the BASTI database\footnote{
{\sf www.oa-teramo.inaf.it/BASTI}}  
with the exact CNO and metallicity (determined in our previous work and 
listed in Tab.~\ref{paramSGBs}), to
investigate the relative age difference among the \sri\ and \spo\ stars.
As demonstrated by M09 (see their Fig.~19), 
isochrones at the same age, and with metallicity different by
0.15 dex (as observed from high resolution spectra),
cannot reproduce the entire size of the split. 
The isochrone fitting accounting for both the metallicity and
CNO variation to the SGB region is shown in Fig.~\ref{iso}  
for the $m_{\rm F606W}$ versus\  $m_{\rm F606W}-m_{\rm F814W}$ CMD. 
The middle blue and red tracks represent the best fitting isochrones to 
the brighter and the fainter SGB respectively.
The age was assumed equal to 13.5 Gyrs which is the value for which the
models give the best fit with data.
For each of the best-fitting tracks we also show 
isochrones with the same chemistry but with the age varied 
by $\pm$300 Myrs., which is the typical error affecting the 
determination of relative ages from isochrone fitting.
It is clear that, by taking into account the observed difference in the CNO
total content, the size of the SGB split is consistent with isochrones 
of the same age.
Note that a possible He variation (if any), in this range of
metallicity and ages, is not expected to change the separation in 
luminosity between the two SGBs, but only modify the SGB shape 
(Ventura \etal\ 2009).\nocite{ven09}
From this analysis, the \sri\ stars appear to be coeval (or 
possibly slightly younger) than the \spo\ stars, indicating that
star formation in M22 could have developed very rapidly.
Possible age differences smaller than $\sim$300~Myrs cannot be
distinguished by our results.

%%%%%%%%%%%%%%%%%%%%%%%%%%%%%%%%%%%%%%%%%%%%%%%%%%%%%%%%%%%%%%%%%%%%%%%%%%
\section{Formation scenarios\label{enri}}
%%%%%%%%%%%%%%%%%%%%%%%%%%%%%%%%%%%%%%%%%%%%%%%%%%%%%%%%%%%%%%%%%%%%%%%%%%

In the attempt to understand the chemical and photometric observations
in M22, two different scenarios can be basically explored, depending
on how different bursts of star formation could have occurred,
spatially or temporally separated.

In the first case we may ``simply'' assume star formation bursts occurred 
in separated regions that eventually merged together. 
Then, the \spo\ and \sri\ stars could have been formed out of interstellar 
mediums with slightly different metallicities, and differences in the total
 C+N+O and $s$-process elements.

If instead, the different bursts of star formation have occurred at
different epochs of the cluster evolution, multiple stellar populations 
in M22 are assumed to be due to self-pollution,
with the \sri\ (metal-richer) stars being the younger population,
as they show signatures of pollution from neutron-capture material.
As far as we know, the self-pollution hypothesis in M22 would require 
a number of complex and ``fine-tuned'' assumptions to reconcile 
the stellar yields and the lifetimes of  possible polluters with the
observations (see Sect.~\ref{chem_nuc} for more details).
Under this scenario, any attempt to identify the responsible polluters 
should take into account the rapid evolution of the cluster, that we 
have argued on the basis of observational spectroscopic and photometric 
results coupled with theoretical isochrones.

\subsection{Chemical enrichment\label{chem_nuc}}

Favorable nucleosynthetic sites for the \spro\ production 
are AGB stars with M$\lesssim$3-4${\rm M_{\odot}}$, which are predicted to 
experience multiple third dredge-up events (TDU).
The number of TDUs decreases for lower mass AGB stars until a minimum mass 
for which the conditions for the activation of the TDU are never reached.
 This minimum mass is an increasing function of metallicity 
(\eg, Straniero \etal\ 2003\nocite{str03}).
AGB stars more massive 3-4${\rm M_{\odot}}$ do not experience many 
third dredge-up events.
They are predicted to pollute the intra-cluster medium only in
the $p$-capture products and, for this reason, have been proposed to
be candidate polluters responsible for the light-element
(anti)correlations typical of GCs (D'Antona \& Caloi 2004\nocite{dan04}).
The evolutionary timescale of the less massive AGBs (of order
some hundreds of Myrs; see Tab.~1 in Ventura \etal\ 2009\nocite{ven09}), 
may be consistent with the uncertainty associated to the null age difference 
obtained from our isochrones constructed with different CNO contents.
However, this still leaves problems in completely accounting for
the chemical enrichment history of M22.

A primary difficulty is the bimodal metallicity distribution of M22, with 
the \sri\ stars having higher Fe abundances (M09).
This suggests that M22 has also undergone enrichment in metallicity, 
and hence has been likely polluted by yields from 
Supernovae Type~II (SN~II, M09).

Moreover, in M22 each $s$-group individually defines a Na-O 
anticorrelation, suggesting that both have suffered further enrichment 
from other polluters, \eg, intermediate-mass AGBs, fast-rotating 
massive stars, and/or massive binaries.
Each of the two M22 $s$-groups shows its own second-generation enriched 
in Na and depleted in O (see Fig.~14 in M11a), just like the
patterns seen in all the GCs studied so far with sufficient statistics 
(\eg\ Ram{\'{\i}}rez \& Cohen 2002; Carretta \etal\ 2009\nocite{car09}). 
This chemical pattern is difficult to understand.  
We may speculate that the first stars to form after the first
population of metal-poor, \spo-poor and Na-poor stars, could have been 
the more metal-rich and \sri\ Na-poor stars, as tentatively suggested 
by Marino \etal\ (2012)\nocite{mar12} to interpret
observations in $\omega$~Cen.
These two populations could form, at later times, their own Na-O
anticorrelation. 
A similar scenario would require an efficient star formation and the 
consume of the available intra-cluster material in order to prevent each
stellar burst to be contaminated by the preceding one.

A chemical pattern similar to the one observed in M22, 
even if much more complex, is seen in $\omega$~Cen 
(Johnson \& Pilachowski 2010\nocite{jp10}, 
Marino \etal\ 2011b\nocite{mar11b}; see \S\ref{introduction} 
for more details).
For this most peculiar GC D'Antona \etal\ (2011),\nocite{dan11} 
compared the Na and O abundances with theoretical AGB yields 
to account for the presence of the Na-O anticorrelation at different 
metallicities. 
They envisaged a chemical evolutionary scenario in which, due to the 
large mass of $\omega$~Cen, the material ejected by SN~II
could survive in a torus that collapses back onto the cluster after 
the SN II epoch (see also D'Ercole \etal\ 2008\nocite{derc2008}).
The 3D hydro simulations by Marcolini \etal\ (2006)\nocite{mar06} 
show in fact that the collapse back includes the matter enriched by 
the SN II ejecta.
A similar scenario could be tentatively extended to M22.

Alternatively, as observations in GCs strongly suggest that the 
increase in \spro\ elements is linked to an increase in Fe, it would be 
tempting to speculate that \spro\ elements and Fe may have been produced 
by the same polluters. 
Indeed, as outlined in D'Antona \etal\ (2011),\nocite{dan11} 
there are other possible sites of $s$-nucleosynthesis that have not been
explored well at present, \eg\ the carbon burning shells of lower-mass 
progenitors of SN~II (\eg, The \etal\ 2007\nocite{the07}).
The pollution from these objects may peculiarly become apparent in the 
evolution of the progenitor systems of $\omega$~Cen and M22.

Interestingly, Roederer \etal\ (2011) noted that the \spro\ abundances in the
\sri\ stars could be more consistent with predictions for more massive
AGB stars, those capable of activating the
$\phantom{}^{22}$Ne($\alpha$,$n$)$\phantom{}^{25}$Mg reaction. 
The main difficulty of this scenario is the lack of a correlation 
between \spro\ enrichment and Na within the two groups, which would be 
expected if these elements are all produced by the same AGB stars 
of higher masses.
However, stars with initial masses $>$3~M$_{\odot}$ will evolve in
$\lesssim$300 Myr, which would agree with our derived upper limit
on the age difference of the two SGBs in M22.  
We note here that the efficiency of $s$-processes depends not just on
the number of neutrons but also on the neutron-to-seed nuclei 
(most likely Fe) ratio.
However, the lack of complete grids of theoretical yields for various 
masses at the exact metallicity of M22 (Cristallo \etal\ 2011\nocite{cri11}) 
makes difficult to draw definitive conclusions on the mass of the AGB 
polluters, in the hypothesis that AGBs are effectively the producers of 
the extra $s$-material in M22 \sri\ stars.
More generally, we admit that the proposed self-pollution 
channels require an uncomfortable level fine-tunings.
The uncertainties, and in some cases the lack, of predicted 
theoretical yields from different kinds of polluters, introduces further 
difficulties in interpreting data in the self-pollution framework.

Some of the difficulties encountered by the self-pollution scenario may 
be overcome invoking a spatial separation (instead than a temporal one) 
for the different bursts of star formation in a merger scenario.
Very recently, dynamical simulations by Bekki \& Yong (2011)\nocite{bek11} 
have shown that it is dynamically plausible that two GCs can merge and form a new GC in the
central region of its host dwarf galaxy. 
The host dwarf galaxy is cannibalized through tidal interactions with 
the Milky Way and only the compact nucleus survives as a present-day AGC. 
This scenario has the advantage in explaining in a simple manner
the chemical features of the two $s$-groups of stars in M22 
(\eg, their different metallicity and \spro\ elements content, 
and the presence of an individual Na-O anticorrelation). 
However, one still has to understand why the multiple stellar 
population phenomenon looks similar in all the AGC investigated so far. 
As an example, both in M22 and NGC~1851 the \sri\ stars are slightly
enriched in metallicity. 
In addition if we consider M22 and $\omega$~Cen, in both of those clusters 
the metal-richer and $s$-richer stars populate fainter SGBs. 
The occurrence of these similarities may be more easily understood 
in a self-enrichment scenario.
However, note for completeness  that the distribution on \sri\ and
\spo\ stars on the SGBs of NGC~1851 could be inverted with respect to
M22, as suggested by Carretta \etal\ (2011)\nocite{car11}, but no direct
observations of SGB are currently available for this GC.

At present we are unable to provide a definitive
and clear explanation for the M22's formation and evolution.
Both the self-enrichment and the merger scenarios have pro's and con's.
We only note here that they represent a very different way to interpret
 M22 and other AGCs. 
In the first hypothesis they could be similar to \textit{normal} clusters, 
but the star formation could have occurred further, and/or owing 
to their initial higher masses could have retained material 
escaped from \textit{normal} GCs. 
In the second hypothesis the AGCs may have been formed through 
different mechanisms (i.e. a merger) in ''exceptional'' conditions,
most probably in extra-galactic environments.

\section{Conclusion\label{conclusions}}

We have presented a medium resolution spectroscopic analysis of a 
hundred SGB stars in the double-SGB GC M22.
The faint SGB is populated by \sri\ (metal-rich) stars, and the 
bright SGB by \spo\ (metal-poor) stars.
Our abundance analysis constitutes the first direct evidence
for the connection between the two RGBs populated by the two \spro\ 
stellar groups, discovered in our previous work, and the double SGB.
This SGBs-RGBs connection has also been confirmed by the inspection
of the $U$-$(U-V)$ CMD: the fainter SGB population clearly
{\it evolves} in a redder RGB sequence populated by \sri\
stars, and the brighter SGB in a bluer branch populated instead
by the \spo\ stars (M11a).
Among the  RGB stars studied by M11a, the \sri\ stars are
also enhanced in the total CNO, and have a mean higher C with respect
to the \spo\ ones.
We can at least extend their conclusion about C and the total CNO to the present SGB sample.

Isochrones constructed with our observed metallicities and C+N+O content
for the two $s$ groups observed on the RGB suggest that the
split SGB is 
consistent with the two stellar groups being coeval within an uncertainty 
of $\sim$300 Myrs.

Based on our observations, we discussed possible evolutionary histories for the
cluster, both in a self-enrichment and in a merger scenario.
We underline the difficulties that both scenarios have to overcome and
we encourage further investigations on both the theoretical and
observational side to finally shed light on the nature of the Milky
Way AGCs.

\begin{acknowledgements}
We thank F. D'Antona, R. Gratton and C. Allende Prieto for useful 
comments on the manuscript, and the anonymous referee for his/her suggestions.
APM, GP, SC and AA are funded by the Ministry of Science and
Technology of the Kingdom of Spain (grant AYA 2010-16717). 
APM and AA are also funded by the Instituto de Astrofísica de Canarias 
(grant P3-94).
IUR is supported by the Carnegie Institution of Washington through 
the Carnegie Observatories Fellowship.
CS is funded with U.S. National Science Foundation grant AST-0908978.
MZ acknowledges the FONDAP Center for Astrophysics 15010003,
the BASAL CATA PFB-06, the Milky Way Millennium Nucleus from the
Ministry of Economics ICM grant P07-021-F,  Proyecto FONDECYT
Regular 1110393, and Proyecto Anillo ACT-86
CS
\end{acknowledgements}

%
%--------------------------------------------------------------------
%

\bibliographystyle{aa}

%----------------------------------------------------------------
\begin{table*}[ht!]
\begin{center}  
\caption{Ground-based photometric database.}
\begin{tabular}{lll}
\hline
\hline
Telescope       &Dates            & Camera               \\
\hline
ESO Dutch 0.91m & 1997 Apr 12-16  & CCD Tektronix        \\
CTIO 0.9m       & 1998 Apr 16-22  & CCD ${\rm Tek2K\_3}$ \\ 
CTIO 0.9m       & 1991 Sep 18-29  & CCD 772              \\
JKT 1.0m        & 1998 Jun 20-26  & CCD TEK4             \\
ESO NTT 3.6m    & 1993 Jul 15-23  & EMMI+Tektronix       \\
ESO NTT 3.6m    & 1993 May 30-31  & SUSI2+EEV44-80        \\
ESO/MPI 2.2m    & 2002 Jun 17-21  & WFI                  \\
ESO/MPI 2.2m    & 1999 May 12-15  & WFI                  \\
ESO/MPI 2.2m    & 2000 Jul 06-12  & WFI                  \\
ESO/MPI 2.2m    & 1999 Jul 06 12  & WFI                  \\
ESO/MPI 2.2m    & 2004 Jun 13-28  & WFI                  \\
\hline
\label{stetson}
\end{tabular}
\end{center}
\end{table*}
%----------------------------------------------------------------

%----------------------------------------------------------------

\longtab{2}{
\begin{longtable}{rccccccr}
\caption{M 22 targets: positions, and photometric data.\label{phot_data_tab}}\\
\hline\hline
ID \tablefootmark{a}  & $\alpha$(2000) & $\delta$(2000)  &$U$\tablefootmark{b}   &$B$\tablefootmark{c}    &$V$\tablefootmark{c}&$I$\tablefootmark{c} & $\Delta~E(B-V)$\tablefootmark{d}\\
\hline
\endfirsthead
\caption{continued.}\\
\hline\hline
ID \tablefootmark{a}  & $\alpha$(2000) & $\delta$(2000) &$U$\tablefootmark{b}&$B$\tablefootmark{c}    &$V$\tablefootmark{c}&$I$\tablefootmark{c} & $\Delta~E(B-V)$\tablefootmark{d}\\
\hline
\endhead
\endfoot
  3&	  18:35:45.23&      $-$23:52:23.7 &     99.999  &	18.016 &	17.159 &	16.049 &	 $-$0.004         \\
   24&	  18:35:53.65&      $-$23:58:04.5 &     99.999  &	17.943 &	17.125 &	16.053 &	 $-$0.013         \\
   77&	  18:36:01.27&      $-$23:55:38.0 &     99.999  &	17.845 &	16.976 &	15.839 &	 $-$0.015         \\
   94&	  18:36:02.73&      $-$23:53:24.6 &     99.999  &	18.207 &	17.330 &	16.170 &	 $-$0.004         \\
   99&	  18:36:03.12&      $-$23:52:33.1 &     99.999  &	18.082 &	17.249 &	15.983 &	       0.001         \\
  112&	  18:36:04.09&      $-$23:49:31.0 &     99.999  &	18.091 &	17.235 &	16.129 &	       0.007         \\
  151&	  18:36:06.55&      $-$23:55:03.2 &     99.999  & 	17.937 &	17.102 &	16.029 &	 $-$0.019         \\
  221&	  18:36:10.54&      $-$23:56:42.5 &     99.999  & 	17.981 &	17.116 &	15.969 &	 $-$0.012         \\
  254&	  18:36:12.08&      $-$23:51:11.8 &     99.999  &	18.053 &	17.225 &	16.114 &	       0.012         \\
  262&	  18:36:12.62&      $-$23:50:02.4 &     99.999  & 	17.998 &	17.160 &	16.044 &	 $-$0.006         \\
  263&	  18:36:12.59&      $-$23:55:48.8 &     19.689  & 	17.875 &	16.916 &	15.704 &	 $-$0.016         \\
  268&	  18:36:12.79&      $-$23:52:57.5 &     19.857  & 	18.130 &	17.247 &	16.089 &	 $-$0.005         \\
  277&	  18:36:12.95&      $-$23:56:02.8 &     19.743  & 	17.943 &	17.046 &	15.861 &	 $-$0.016         \\
  290&	  18:36:13.28&      $-$23:52:31.0 &     19.773  &	17.983 &	17.035 &	15.822 &	       0.007         \\
  292&	  18:36:13.36&      $-$23:53:57.7 &     99.999  & 	17.756 &	16.873 &	15.738 &	 $-$0.015         \\
  298&	  18:36:13.56&      $-$23:54:27.0 &     19.615  &	17.912 &	17.064 &	15.984 &	       0.002         \\
  312&	  18:36:14.05&      $-$23:52:21.4 &     19.388  &	17.656 &	16.700 &	15.466 &	       0.003         \\
  329&	  18:36:14.48&      $-$23:54:58.0 &     99.999  &	17.842 &	16.967 &	15.839 &	       0.006         \\
  351&	  18:36:15.08&      $-$23:57:01.1 &     19.907  & 	18.049 &	17.159 &	16.002 &	 $-$0.017         \\
  352&	  18:36:15.08&      $-$23:56:41.3 &     20.048  & 	18.263 &	17.424 &	16.339 &	 $-$0.006         \\ 
  374&	  18:36:15.60&      $-$23:58:31.5 &     99.999  & 	17.796 &	16.913 &	15.737 &	 $-$0.043         \\
  375&	  18:36:15.62&      $-$23:55:28.1 &     20.054  &	18.299 &	17.439 &	16.227 &	       0.004         \\
  398&	  18:36:16.25&      $-$23:52:33.1 &     19.538  & 	17.844 &	16.905 &	15.704 &	 $-$0.009         \\
  456&	  18:36:17.09&      $-$23:54:07.9 &     19.567  & 	17.675 &	16.787 &	15.624 &	 $-$0.001         \\
  472&	  18:36:17.48&      $-$23:54:45.0 &     20.040  &	18.267 &	17.448 &	16.391 &	       0.011         \\
  479&	  18:36:17.53&      $-$23:50:30.3 &     99.999  & 	18.154 &	17.280 &	16.140 &	 $-$0.004         \\
  494&	  18:36:17.73&      $-$23:49:41.6 &     99.999  &	18.129 &	17.237 &	16.094 &	       0.012         \\
  557&	  18:36:18.40&      $-$23:53:04.0 &     20.021  &	18.167 &	17.330 &	16.192 &	       0.007         \\
  575&	  18:36:18.60&      $-$23:54:30.1 &     99.999  &	18.416 &	17.613 &	16.572 &	       0.009         \\ 
  614&	  18:36:18.93&      $-$23:52:19.1 &    19.784   & 	18.035 &	17.111 &	15.923 &	 $-$0.013         \\
  718&	  18:36:19.90&      $-$23:53:55.4 &     99.999  &	17.543 &	16.735 &	15.660 &	       0.011         \\
  725&	  18:36:19.96&      $-$23:54:34.8 &     20.168  &	18.546 &	17.756 &	16.732 &	       0.014         \\
  729&	  18:36:20.00&      $-$23:53:00.5 &     20.079  &	18.295 &	17.424 &	16.264 &	       0.009         \\
  735&	  18:36:20.03&      $-$23:56:28.4 &     19.602  & 	17.883 &	17.024 &	15.900 &	 $-$0.001         \\
  768&	  18:36:20.25&      $-$23:56:01.4 &     19.684  &	17.934 &	17.046 &	15.897 &	       0.004         \\
  810&	  18:36:20.53&      $-$23:55:28.2 &     99.999  &	18.079 &	17.303 &	15.738 &	       0.013         \\
  911&	  18:36:21.19&      $-$23:53:32.6 &     99.999  &	17.974 &	17.175 &	16.149 &	       0.014         \\
  972&	  18:36:21.56&      $-$23:53:48.4 &     99.999  &	17.574 &	16.724 &	15.633 &	       0.003         \\
  975&	  18:36:21.56&      $-$23:56:46.8 &    19.864   &	18.116 &	17.259 &	16.131 &	       0.001         \\
  993&	  18:36:21.69&      $-$23:52:30.8 &     20.137  &	18.486 &	17.664 &	16.568 &	       0.020         \\
 1037&	  18:36:21.93&      $-$23:55:38.6 &     19.589  & 	17.891 &	17.040 &	15.876 &	       0.015         \\
 1110&	  18:36:22.40&      $-$23:52:25.6 &     19.578  & 	17.823 &	16.919 &	15.758 &	       0.026         \\
 1114&	  18:36:22.41&      $-$23:54:37.2 &     99.999  & 	18.548 &	17.727 &	16.644 &	       0.009         \\
 1116&	  18:36:22.42&      $-$23:53:41.5 &     99.999  & 	17.951 &	17.017 &	15.795 &	       0.001         \\
 1138&	  18:36:22.58&      $-$23:57:08.0 &     99.999  &	17.847 &	16.904 &	15.721 &	       0.011         \\
 1257&	  18:36:23.35&      $-$23:55:33.9 &     99.999  &	18.212 &	17.396 &	16.320 &	       0.021         \\
 1289&      18:36:23.59 &     $-$23:52:38.9 &     19.680  &    18.019  & 17.141 &  16.041&         0.024         \\
 1451&	  18:36:24.56&      $-$23:54:56.1 &     99.999  & 	18.267 &	17.411 &	16.287 &	       0.015         \\
 1526&	  18:36:24.98&      $-$23:52:37.8 &     99.999  & 	18.058 &	17.243 &	16.167 &	       0.018         \\
 1637&	  18:36:25.65&      $-$23:54:31.7 &     99.999  & 	18.351 &	17.379 &	16.323 &	       0.020         \\
 1663&	  18:36:25.84&      $-$23:52:35.9 &     99.999  & 	18.032 &	17.184 &	16.083 &	       0.013         \\
 1685&	  18:36:26.06&      $-$23:54:21.2 &     99.999  &	17.710 &	16.871 &	15.774 &	       0.018         \\
 1763&	  18:36:26.68&      $-$23:54:59.2 &     99.999  &	18.346 &	17.525 &	16.404 &	       0.019         \\
 1792&	  18:36:26.86&      $-$23:54:36.4 &     99.999  &	18.102 &	17.200 &	16.178 &	       0.016         \\
 1821&      18:36:27.09&      $-$23:52:14.3 &     99.999  &   18.203 & 17.314 &  16.177 &        0.010           \\
 1894&	  18:36:27.53&      $-$23:53:53.2 &     99.999  &  	18.062 &	17.219 &	16.117 &	       0.017         \\ 
 1903&	  18:36:27.61&      $-$23:58:00.7 &     99.999  &  	17.924 &	17.125 &	16.039 &	 $-$0.030         \\
 1913&	  18:36:27.69&      $-$23:55:08.4 &     99.999  &	17.840 &	16.941 &	15.797 &	       0.025         \\
 1924&	  18:36:27.74&      $-$23:53:19.0 &     99.999  &	17.975 &	17.088 &	15.967 &	       0.010         \\ 
 1939&	  18:36:27.85&      $-$23:54:12.7 &     99.999  &	17.920 &	17.053 &	15.982 &	       0.013         \\
 1988&	  18:36:28.26&      $-$23:54:50.2 &     99.999  &	18.276 &	17.464 &	16.340 &	       0.022         \\
 1993&	  18:36:28.29&      $-$23:55:40.3 &     99.999  &	18.075 &	17.146 &	15.937 &	       0.010         \\
 2061&	  18:36:29.01&      $-$23:53:35.4 &     99.999  &	17.905 &	17.139 &	16.071 &	       0.016         \\
 2099&	  18:36:29.36&      $-$23:56:09.2 &     99.999  & 	18.140 &	17.274 &	16.152 &	       0.001         \\
 2153&	  18:36:29.78&      $-$23:52:55.4 &     99.999  &  	18.103 &	17.238 &	16.141 &	 $-$0.006         \\
 2161&	  18:36:29.88&      $-$23:59:04.3 &     99.999  &  	17.965 &	17.085 &	15.964 &	 $-$0.046         \\
 2175&	  18:36:30.02&      $-$23:55:48.1 &     99.999  &	18.173 &	17.280 &	16.131 &	       0.012         \\
 2201&	  18:36:30.25&      $-$23:52:25.5 &     99.999  &  	18.043 &	17.175 &	16.083 &	 $-$0.003         \\
 2207&	  18:36:30.28&      $-$23:50:34.8 &     99.999  &	17.808 &	16.933 &	15.778 &	       0.006         \\
 2209&	  18:36:30.28&      $-$23:55:33.5 &     99.999  &	18.224 &	17.391 &	16.298 &	       0.014         \\ 
 2242&	  18:36:30.65&      $-$23:58:20.9 &     99.999  &  	17.978 &	17.178 &	16.073 &	 $-$0.033         \\
 2278&	  18:36:30.98&      $-$23:53:59.7 &     99.999  &	17.888 &	17.049 &	15.968 &	       0.003         \\
 2300&	  18:36:31.14&      $-$23:52:20.6 &     99.999  &  	17.999 &	17.206 &	16.150 &	 $-$0.003         \\
 2312&	  18:36:31.32&      $-$23:52:58.3 &     99.999  &	17.776 &	16.931 &	15.882 &	       0.002         \\
 2318&	  18:36:31.35&      $-$23:53:47.8 &     99.999  &	17.979 &	17.140 &	16.085 &	       0.008         \\
 2321&	  18:36:31.41&      $-$23:57:44.3 &     99.999  &  	17.664 &	16.777 &	15.621 &	 $-$0.019         \\
 2334&	  18:36:31.55&      $-$23:53:12.0 &     99.999  & 	17.980 &	17.066 &	15.937 &	       0.006         \\
 2353&	  18:36:31.69&      $-$23:53:25.8 &     99.999  &	18.150 &	17.380 &	16.373 &	       0.006         \\
 2364&	  18:36:31.92&      $-$24:00:17.7 &     99.999  &  	17.860 &	16.960 &	15.792 &	 $-$0.039         \\
 2391&	  18:36:32.50&      $-$23:55:37.5 &     99.999  &	17.917 &	16.999 &	15.796 &	        0.020         \\
 2404&	  18:36:32.70&      $-$24:01:01.7 &     99.999  &  	18.087 &	17.257 &	16.152 &	 $-$0.036         \\ 
 2414&	  18:36:33.06&      $-$23:57:10.2 &     99.999  & 	18.165 &	17.294 &	16.189 &	       0.001         \\
 2419&	  18:36:33.20&      $-$23:57:58.6 &     99.999  &  	17.985 &	17.141 &	16.058 &	 $-$0.021         \\
 2421&	  18:36:33.27&      $-$23:50:13.8 &     99.999  &  	18.108 &	17.252 &	16.065 &	 $-$0.000         \\ 
 2481&	  18:36:34.86&      $-$23:50:30.7 &     99.999  &  	18.039 &	17.199 &	16.090 &	 $-$0.000         \\ 
 2505&	  18:36:35.65&      $-$23:57:57.9 &     99.999  &  	18.136 &	17.303 &	16.210 &	 $-$0.012         \\ 
 2542&	  18:36:36.81&      $-$23:55:46.6 &     99.999  & 	18.041 &	17.105 &	15.882 &	       0.019         \\ 
 2544&	  18:36:36.85&      $-$23:50:25.1 &     99.999  &  	18.004 &	17.137 &	15.998 &	 $-$0.002         \\ 
 2570&	  18:36:38.23&      $-$23:51:12.5 &     99.999  &  	17.961 &	17.133 &	16.012 &	 $-$0.005         \\
 2572&	  18:36:38.33&      $-$23:52:51.0 &     99.999  &	18.001 &	17.159 &	16.065 &	       0.004         \\
 2590&	  18:36:39.01&      $-$23:58:07.9 &     99.999  & 	18.043 &	17.113 &	15.888 &	       0.003         \\
 2607&	  18:36:39.89&      $-$23:54:02.5 &     99.999  & 	18.264 &	17.393 &	16.267 &	       0.014         \\
 2621&	  18:36:40.64&      $-$23:54:20.6 &     99.999  &  	18.030 &	17.090 &	15.884 &	       0.009         \\
 2625&	  18:36:41.07&      $-$23:51:22.1 &     99.999  &  	18.049 &	17.228 &	16.109 &	 $-$0.018         \\
 2650&	  18:36:42.62&      $-$23:52:49.9 &     99.999  &	18.058 &	17.141 &	15.996 &	       0.012         \\
 2659&	  18:36:43.23&      $-$23:53:19.0 &     99.999  &	18.080 &	17.207 &	16.078 &	       0.012         \\
 2669&	  18:36:44.22&      $-$23:48:31.5 &     99.999  &  	17.976 &	17.073 &	15.897 &	 $-$0.001         \\
 2672&	  18:36:44.63&      $-$23:49:45.3 &     99.999  &	17.783 &	16.874 &	15.697 &	       0.012         \\
 2689&	  18:36:45.66&      $-$23:54:21.6 &     99.999  &	18.274 &	17.377 &	16.209 &	       0.015         \\
 2801&	  18:36:59.22&      $-$23:52:16.7 &     99.999  &  	17.913 &	17.049 &	15.914 &	 $-$0.011         \\
 2815&	  18:37:03.73&      $-$23:49:56.7 &     99.999  &  	17.932 &	17.009 &	15.824 &	 $-$0.004         \\
\hline
\end{longtable}
\tablefoottext{a}{Identification numbers come from the ground-based
  photometric catalog described in Sect.~\ref{sect_phot}}
\tablefoottext{b}{Momany et al. (2004) photometric data-base.}
\tablefoottext{c}{Stetson photometric data-base.}
\tablefoottext{d}{Differential reddening correction values for each target star.}
}

%----------------------------------------------------------------

%----------------------------------------------------------------
\longtab{3}{
\begin{longtable}{rccrrrrl}
\caption{M22 C, Sr, Ba abundances for SGB stars. The standard
  deviation $\sigma_{\rm Sr}$  for the two Sr lines and the $s$-group
  are also listed for each star.\label{abb_data_tab}}\\
\hline\hline
ID  & \teff\ & \logg\ &\phantom{}[Sr/Fe]& \phantom{}$\sigma_{\rm Sr}$& \phantom{}[Ba/Fe] & \phantom{}[C/Fe]&$s$-group \tablefootmark{a}  \\
\hline
\endfirsthead
\caption{continued.}\\
\hline\hline
ID  & \teff\ & \logg\ &\phantom{}[Sr/Fe]& \phantom{}$\sigma_{\rm Sr}$ & \phantom{}[Ba/Fe] & \phantom{}[C/Fe]& $s$-group \tablefootmark{a}  \\
\hline
\endhead
     3 &   5866 &  3.88   &  $-$0.09   &  0.12  &  $-$0.12 &     0.19  &	           \spo\	 \\
    24 &   5942 &  3.92   &   0.06   &  0.14  &   0.09 &    $-$0.18  &               \spo\	 \\ 
    77 &   5774 &  3.78   &   0.04   &  0.21  &   0.00 &    $-$0.22  &               \spo\	 \\ 
    94 &   5794 &  3.92   &   0.40   &  0.11  &   0.46 &    $-$0.09  &               \sri\	 \\ 
    99 &   6026 &  3.96   &  $-$0.19   &  0.07  &  $-$0.19 &    $-$0.20  &               \spo\	 \\ 
   112 &   5917 &  3.92   &  $-$0.00   &  0.13  &   0.03 &    $-$0.40  &               \spo\	 \\
   151 &   5872 &  3.89   &  $-$0.15   &  0.06  &  $-$0.09 &    $-$0.15  &               \spo\	 \\
   221 &   5808 &  3.85   &   0.11   &  0.05  &   0.10 &    $-$0.31  &               \spo\	 \\
   254 &   6030 &  3.96   &  $-$0.12   &  0.03  &  $-$0.21 &     0.00  &               \spo\	 \\
   262 &   5939 &  3.91   &  $-$0.21   &  0.05  &   0.04 &    $-$0.17  &               \spo\    	 \\
   263 &   5472 &  3.62   &   0.56   &  0.00  &   0.60 &     0.25  &               \sri\	 \\
   268 &   5744 &  3.87   &   0.37   &  0.09  &   0.21 &     0.03  &               \sri\	 \\
   277 &   5689 &  3.77   &  $-$0.22   &  0.06  &   0.03 &    $-$0.31  &               \spo\	 \\
   290 &   5552 &  3.69   &   0.38   &  0.11  &   0.46 &     0.20  &               \sri\	 \\
   292 &   5743 &  3.72   &  $-$0.05   &  0.11  &   0.21 &    $-$0.05  &               \spo\	 \\
   298 &   5928 &  3.86   &   0.07   &  0.02  &  $-$0.05 &    $-$0.05  &               \spo\	 \\
   312 &   5514 &  3.54   &  $-$0.12   &  0.01  &   0.07 &    $-$0.12  &               \spo\	 \\
   329 &   5834 &  3.78   &   0.27   &  0.01  &   0.03 &    $-$0.30  &               \sri\	 \\
   351 &   5685 &  3.82   &   0.20   &  0.20  &   0.43 &     0.16  &               \sri\	 \\
   352 &   5911 &  4.01   &   0.29   &  0.04  &   0.14 &    $-$0.14  &               \sri\	 \\
   374 &   5592 &  3.73   &   0.16   & 0.18  &   0.27 &    $-$0.02  &               \spo\	 \\
   375 &   5900 &  3.99   &   0.13   &  0.02  &   0.24 &    $-$0.32  &               \sri\	 \\
   398 &   5558 &  3.65   &  $-$0.12   &  0.01  &  $-$0.08 &    $-$0.12  &               \spo\	 \\
   456 &   5753 &  3.68   &  $-$0.12   &  0.01  &   0.08 &    $-$0.09  &               \spo\	 \\
   472 &   6103 &  4.06   &   0.05   &  0.00  &  $-$0.26 &     0.05  &               \spo\	 \\ 
   479 &   5792 &  3.90   &   0.21   &  0.01  &   0.16 &    $-$0.12  &               \sri\	 \\ 
   494 &   5796 &  3.86   &   0.22   &  0.18  &   0.13 &     0.20  &               \sri\	 \\ 
   557 &   6002 &  3.98   &   0.36   &  0.12  &   0.46 &     0.25  &               \sri\	 \\ 
   575 &   6236 &  4.16   &   0.50   &  0.04  &   0.27 &     0.40  &               \sri\	 \\ 
   614 &   5589 &  3.75   &   0.45   &  0.04  &   0.37 &    $-$0.08  &               \sri\	 \\ 
   718 &   6134 &  3.79   &   0.09   &  0.16  &   0.02 &     0.22  &               \sri\	 \\ 
   725 &   6301 &  4.24   &  $-$0.16   &  0.07  &  $-$0.23 &     0.00  &               \spo\	 \\
   729 &   5864 &  3.97   &   0.32   &  0.04  &   0.13 &     0.03  &               \sri\	 \\ 
   735 &   5874 &  3.82   &  $-$0.01   &  0.08  &   0.08 &    $-$0.28  &               \spo\	 \\
   768 &   5759 &  3.79   &  $-$0.05   &  0.04  &   0.12 &     0.00  &               \spo\	 \\ 
   810 &   6311 &  4.08   &   0.32   &  0.07  &   0.87 &     0.52  &               \sri\	 \\ 
   911 &   6216 &  3.99   &  $-$0.15   &  0.00  &  $-$0.09 &     0.40  &               \spo\	 \\
   972 &   5949 &  3.72   &   0.01   &  0.08  &  $-$0.12 &     0.22  &               \spo\	 \\ 
   975 &   5890 &  3.92   &   0.32   &  0.08  &   0.33 &     0.21  &               \sri\	 \\ 
   993 &   6112 &  4.15   &   0.16   &  0.04  &   0.43 &     0.13  &               \sri\	 \\ 
  1037 &   5980 &  3.85   &  $-$0.09   &  0.01  &   0.11 &     0.14  &               \spo\	 \\ 
  1110 &   5800 &  3.72   &   0.01   &  0.23  &   0.01 &    $-$0.15  &               \spo\	 \\ 
  1114 &   6113 &  4.17   &   0.26   &  0.03  &   0.15 &     0.32  &               \sri\	 \\ 
  1116 &   5639 &  3.71   &  $-$0.15   &  0.04  &  $-$0.01 &    $-$0.19  &               \spo\	 \\
  1138 &   5612 &  3.65   &   0.41   &  0.14  &   0.41 &     0.06  &               \sri\	 \\ 
  1257 &   6162 &  4.05   &  $-$0.20   &  0.07  &  $-$0.23 &     0.20  &  	           \spo\	 \\
  1289 &   5781 &  3.81   &   0.06   &  0.05  &  $-$0.03 &    $-$0.12  &  		   \spo\	 \\
  1451 &   5950 &  3.99   &   0.35   &  0.09  &   0.59 &     0.18  &  		   \sri\	 \\ 
  1526 &   6141 &  3.99   &  $-$0.29   &  0.09  &   0.34 &     0.31  &  		   \spo\	 \\ 
  1637 &   5539 &  3.80   &  $-$0.14   &  0.04  &   0.09 &    $-$0.43  &  		   \spo\	 \\
  1663 &   5988 &  3.91   &  $-$0.08   &  0.08  &   0.18 &     0.09  &  		   \spo\	 \\ 
  1685 &   6051 &  3.80   &   0.18   &  0.12  &   0.00 &     0.06  &  		   \sri\	 \\ 
  1763 &   6072 &  4.09   &  $-$0.18   &  0.05  &  $-$0.15 &     0.00  &  		   \spo\	 \\
  1792 &   5783 &  3.83   &  $-$0.12   &  0.23  &  $-$0.32 &    $-$0.15  &  		   \spo\	 \\ 
  1821 &   5824 &  3.90   &   0.21   &  ...  &   ... &     ...  &  	 	   \sri\	 \\ 
  1894 &   6074 &  3.94   &   0.38   &  0.06  &   0.56 &     0.64  &               \sri\	 \\ 
  1903 &   5975 &  3.95   &   0.09   &  0.11  &  $-$0.28 &    $-$0.11  &               \spo\	 \\
  1913 &   5810 &  3.73   &  $-$0.02   &  0.16  &   0.11 &    $-$0.08  &               \spo\	 \\
  1924 &   5856 &  3.81   &   0.31   &  0.11  &   0.47 &    $-$0.06  &               \sri\	 \\ 
  1939 &   5917 &  3.83   &  $-$0.22   &  0.11  &  $-$0.11 &    $-$0.11  &               \spo\	 \\
  1988 &   6209 &  4.09   &   0.12   &  0.20  &   0.26 &     0.28  &               \sri\	 \\ 
  1993 &   5691 &  3.77   &   0.47   &  0.02  &   0.37 &    $-$0.06  &               \sri\	 \\ 
  2061 &   6361 &  4.03   &   0.27   &  0.00  &  $-$0.16 &     0.69  &               \sri\	 \\ 
  2099 &   5862 &  3.91   &   0.27   &  0.10  &   0.51 &     0.23  &               \sri\	 \\ 
  2153 &   5840 &  3.90   &   0.20   &  0.15  &   0.21 &     0.07  &               \sri\	 \\ 
  2161 &   5612 &  3.80   &   0.38   &  0.07  &   0.76 &    $-$0.15  &               \sri\	 \\ 
  2175 &   5795 &  3.88   &   0.30   &  0.23  &   0.04 &     0.36  &               \sri\	 \\ 
  2201 &   5833 &  3.87   &   0.17   &  0.23  &   0.07 &     0.01  &               \spo\	 \\ 
  2207 &   5819 &  3.76   &   0.04   &  0.20  &  $-$0.20 &     0.00  &               \spo\	 \\ 
  2209 &   6061 &  4.02   &   0.12   &  0.08  &   0.09 &     0.20  &               \spo\	 \\ 
  2242 &   5959 &  3.97   &   0.50   &  0.04  &   0.38 &     0.03  &               \sri\	 \\ 
  2278 &   5975 &  3.87   &  $-$0.18   &  0.08  &  $-$0.22 &     0.04  &               \spo\	 \\
  2300 &   6158 &  4.00   &  $-$0.02   &  0.01  &  $-$0.17 &     0.10  &               \spo\	 \\
  2312 &   5976 &  3.81   &   0.17   &  0.19  &   0.24 &     0.11  &               \sri\	 \\ 
  2318 &   5990 &  3.90   &  $-$0.08   &  0.06  &  $-$0.28 &     0.07  &               \spo\	 \\
  2321 &   5675 &  3.67   &  $-$0.19   &  0.05  &  $-$0.20 &    $-$0.24  &               \spo\	 \\ 
  2334 &   5704 &  3.76   &   0.34   &  0.10  &   0.19 &     0.03  &               \sri\	 \\ 
  2353 &   6307 &  4.12   &  $-$0.04   &  0.03  &   0.14 &     0.22  &               \spo\	 \\ 
  2364 &   5569 &  3.72   &  $-$0.09   &  0.04  &   0.00 &    $-$0.19  &               \spo\	 \\
  2391 &   5704 &  3.73   &  $-$0.07   &  0.07  &   0.14 &     0.23  &  	           \spo\	 \\ 
  2404 &   5836 &  3.95   &   0.26   &  0.09  &   0.47 &    $-$0.11  &   		   \sri\	 \\ 
  2414 &   5830 &  3.91   &   0.29   &  0.01  &   0.39 &     0.22  &  		   \sri\	 \\ 
  2419 &   5844 &  3.89   &  $-$0.16   &  0.04  &  $-$0.14 &     0.00  &  		   \spo\	 \\
  2421 &   5902 &  3.92   &   0.22   &  0.18  &   0.33 &     0.14  &  		   \sri\	 \\ 
  2481 &   5948 &  3.92   &   0.02   &  0.22  &   0.12 &    $-$0.40  &  		   \spo\	 \\ 
  2505 &   5923 &  3.97   &   0.30   &  0.01  &   0.15 &     0.16  &  		   \sri\	 \\ 
  2542 &   5656 &  3.74   &  $-$0.05   &  0.01  &   0.23 &    $-$0.24  &  		   \spo\	 \\
  2544 &   5839 &  3.86   &  $-$0.06   &  0.09  &  $-$0.13 &    $-$0.06  &  		   \spo\	 \\ 
  2570 &   5994 &  3.92   &  $-$0.25   &  0.12  &   0.14 &    $-$0.11  &  		   \spo\	 \\
  2572 &   5969 &  3.91   &   0.10   &  0.06  &  $-$0.09 &    $-$0.25  &  		   \spo\	 \\
 2590 &   5599 &  3.75   &  $-$0.08   &  0.05  &  $-$0.19 &    $-$0.55  &   		   \spo\	 \\ 
 2607 &   5872 &  3.96   &   0.31   &  0.01  &   0.16 &     0.27  &   		   \sri\	 \\ 
 2621 &   5621 &  3.73   &   0.19   &  0.12  &   0.67 &    $-$0.18  &   		   \sri\	 \\ 
 2625 &   5961 &  3.96   &   0.19   &  0.14  &   0.54 &    $-$0.01  &   		   \sri\	 \\ 
 2650 &   5666 &  3.78   &   0.15   &  0.21  &   0.13 &    $-$0.05  &   		   \sri\	 \\ 
 2659 &   5863 &  3.88   &   0.38   &  0.04  &   0.53 &    $-$0.04  &   		   \sri\	 \\ 
 2669 &   5721 &  3.78   &  $-$0.04   &  0.12  &   0.18 &    $-$0.03  &   		   \spo\	 \\ 
 2672 &   5721 &  3.69   &  $-$0.09   &  0.04  &  $-$0.11 &    $-$0.15  &   		   \spo\	 \\ 
 2689 &   5826 &  3.91   &   0.19   &  0.01  &   0.11 &    $-$0.27  &   		   \sri\	 \\ 
 2801 &   5814 &  3.82   &  $-$0.07   &  0.03  &  $-$0.23 &    $-$0.28  &   		   \spo\	 \\ 
 2815 &   5635 &  3.72   &   0.08   &  0.03  &  $-$0.08 &     0.03  &   		   \spo\	 \\ 
\hline
\end{longtable}
\tablefoottext{a}{Global metallicities of [A/H]=$-$1.82 and
 [A/H]=$-$1.67 were employed for \spo\ and \sri\ stars, respectively
 (see \S~3.1 for more details). }
}

%----------------------------------------------------------------

%----------------------------------------------------------------

\begin{table*}
\scriptsize
\begin{center}  
\caption{Chemical features of the two M22 SGB. The [Fe/H] and the CNO
 total content are the values determined by Marino \etal\ (2011) for
 RGB stars.}
\begin{tabular}{lcccccccccc}
\hline
\hline
                 & [Fe/H]\tablefootmark{a}     &$\sigma$   & [C+N+O/Fe]\tablefootmark{a}  &$\sigma$& [C/Fe] &$\sigma$ &[Sr/Fe] &$\sigma$&[Ba/Fe]&$\sigma$ \\
\hline
SGB-faint   & $-$1.67$\pm$0.01    &0.05 &$+$0.41$\pm$0.02&0.07 &  $+$0.10$\pm$0.03 &0.23&$+$0.29$\pm$0.02  &0.11& $+$0.32$\pm$0.03&0.22\\
SGB-bright & $-$1.82$\pm$0.02    &0.07 &$+$0.28$\pm$0.02&0.09 &  $-$0.07$\pm$0.03 &0.19& $-$0.06$\pm$0.01  &0.11& $-$0.03$\pm$0.02&0.16 \\
\hline
\label{paramSGBs}
\end{tabular}
\end{center}
\tablefoottext{a}{For these two columns values obtained from RGB stars in
 Marino \etal\ (2009, 2011) are listed.}
\end{table*}

%----------------------------------------------------------------

\end{document}